\documentclass[sigplan,nonacm,10pt]{acmart}

\usepackage{algorithmic}
\usepackage{graphicx}
\usepackage[table]{xcolor}
\usepackage{url}
\PassOptionsToPackage{hyphens}{url}\usepackage{hyperref}
\usepackage{listings}
\usepackage{enumitem}
\usepackage[linesnumbered,ruled,noend]{algorithm2e}
\usepackage{subcaption}
\usepackage{multirow}
\usepackage{cleveref}
\usepackage{multicol}
\usepackage{array}
\usepackage{tabularx}

\usepackage{soul}
\usepackage{siunitx}
\usepackage{tikz}
\usepackage{pgfplots}
\pgfplotsset{compat=1.18}
\usepgfplotslibrary{groupplots}
\usetikzlibrary{patterns}
\definecolor{pxblue}{RGB}{0,102,153}
\definecolor{ltorange}{RGB}{230,159,0}
\definecolor{origfill}{RGB}{235,235,235}
\definecolor{good}{HTML}{D3EDD6}
\definecolor{slow}{HTML}{FCE3C4}
\definecolor{ns}{HTML}{F1F1F1}
\newcommand{\takeaway}[1]{%
  \par\smallskip\noindent
  {\setlength{\fboxsep}{4pt}%
   \fcolorbox{black!30}{black!4}{%
    \begin{minipage}{\dimexpr\columnwidth-2\fboxsep-2\fboxrule\relax}%
      \small\textbf{Takeaway.}\ #1%
    \end{minipage}}}%
  \par\smallskip}
\usetikzlibrary{arrows.meta,calc,matrix,fit,backgrounds,patterns,%
                decorations.pathreplacing}

\newcommand{\nnum}[1]{\num{#1}}
\newcommand{\circled}[1]{%
  \tikz[baseline=(char.base)]{
    \node[shape=circle, fill=black, text=white, inner sep=1pt,
          font=\scriptsize] (char) {#1};}}

\SetKw{Continue}{continue}

\setlist[itemize]{noitemsep, leftmargin=1em}
\setlist[enumerate]{noitemsep, leftmargin=1em}

\setlist[itemize]{noitemsep, leftmargin=1em}
\setlist[enumerate]{noitemsep, leftmargin=1em}

\definecolor{dkred}{RGB}{87,10,10}
\definecolor{burntorange}{rgb}{0.8, 0.33, 0.0}
\definecolor{OrangeRed}{rgb}{1.0, 0.27, 0.0}
\definecolor{ForestGreen}{rgb}{0.0, 0.27, 0.13}

\definecolor{groovyblue}{HTML}{0000A0}
\definecolor{dkgreen}{HTML}{008000}
\definecolor{darkgray}{rgb}{.4,.4,.4}
\definecolor{bittersweet}{rgb}{1.0, 0.44, 0.37}

\newcommand{\xtrim}{{\sc PyXtrim}}
\newcommand{\ltrim}{{$\lambda$-trim}}
\newcommand{\slimstart}{{\sc SlimStart}}

\newcommand{\empirical}[1]{#1}

\newcommand{\point}[1]{{\noindent\bf #1:} }

\newcommand{\dedge}{\overset{\texttt{d}}{\rightarrow}}

\newcommand{\shd}[1]{{\normalfont\textsf{#1}}}   
\newcolumntype{Y}{>{\raggedright\arraybackslash}X}

\newcommand{\numberofapps}{31}

\lstdefinelanguage{PythonBytecode}{
  alsoletter={_},
  sensitive=true,
  morekeywords={
    RESUME, NOP, POP_TOP, PUSH_NULL, COPY, SWAP,
    LOAD_CONST, LOAD_NAME, STORE_NAME, DELETE_NAME,
    LOAD_FAST, STORE_FAST, DELETE_FAST, LOAD_FAST_LOAD_FAST,
    LOAD_GLOBAL, STORE_GLOBAL, LOAD_ATTR, STORE_ATTR, LOAD_METHOD,
    IMPORT_NAME, IMPORT_FROM, MAKE_FUNCTION, MAKE_CELL,
    BINARY_OP, BINARY_SUBSCR, STORE_SUBSCR, DELETE_SUBSCR,
    COMPARE_OP, CONTAINS_OP, IS_OP, UNARY_NOT, UNARY_NEGATIVE,
    BUILD_LIST, BUILD_TUPLE, BUILD_MAP, BUILD_SET, BUILD_STRING,
    LIST_APPEND, SET_ADD, MAP_ADD, UNPACK_SEQUENCE,
    CALL, CALL_FUNCTION_EX, KW_NAMES, RETURN_VALUE, RETURN_CONST,
    GET_ITER, FOR_ITER, JUMP_FORWARD, JUMP_BACKWARD,
    POP_JUMP_IF_FALSE, POP_JUMP_IF_TRUE, END_FOR,
  },
  morecomment=[l]{\#},
  morecomment=[l][\color{dkgreen}]{<--},
  morestring=[b]',
}

\begin{document}
\sloppy

\title{Reducing Cold-Start Latency in Serverless Applications via Dynamic Slicing}

\author{%
  \mbox{Georgios Alexopoulos\textsuperscript{1}} \quad
  \mbox{Konstantinos Karakatsanis\textsuperscript{1}} \quad
  \mbox{Nikolaos Alexopoulos\textsuperscript{2}} \quad
  \mbox{Dimitris Mitropoulos\textsuperscript{1}} \quad
  \mbox{Thodoris Sotiropoulos}%
}
\affiliation{%
  \institution{\textsuperscript{1}University of Athens, and National
    Infrastructures for Research and Technology}
  \institution{\textsuperscript{2}Athens University of Economics and
    Business, and National Cybersecurity Authority of Greece}
  \country{}%
}
\email{\{grgalex, konkara, dimitro\}@ba.uoa.gr \quad alexopoulos@aueb.gr \quad theosotir@windowslive.com}
\renewcommand{\shortauthors}{Alexopoulos et al.}

\begin{abstract}
We present~\xtrim,
a system that reduces
the cold-start latency of serverless applications
through debloating.
We focus on Python,
a dominant language for serverless applications
whose dynamic features
and extensive use of native extensions make
traditional static debloating particularly challenging.
\xtrim{} frames debloating
as a~\textit{dynamic slicing} problem,
using the application’s externally visible behavior
as the slicing criterion.
Everything outside the resulting slice is removed,
both from the application and its dependencies.
Our key technical contribution
is that the slice is computed by
a~\textit{cross-language dynamic dependence engine}
that tracks data and control dependences
across Python and native code
and identifies operations that
interact with the operating system,
which form the slicing criterion.
As a result,
\xtrim{} can effectively handle real-world applications
that rely on dynamic features such as reflection,
interoperate with native code and
interact with system resources.
Across~\numberofapps\ applications on AWS Lambda,
\xtrim{} reduces cold-start latency by
\empirical{21.7\%}
and peak memory usage by~\empirical{17.1\%} at the median.
This is more than double the reduction achieved
by the state of the art,
while debloating each application in minutes.
\end{abstract}

\maketitle

\section{Introduction}
\label{sec:intro}
Serverless computing lets developers
run event-driven functions
through entry points called~\textit{handlers},
without provisioning or managing
the underlying servers,
with usage-based billing.
This approach has become a mainstream deployment model,
with platforms such as AWS Lambda,
Azure Functions,
and Google Cloud Functions widely used
in production~\cite{datadog-2025}.

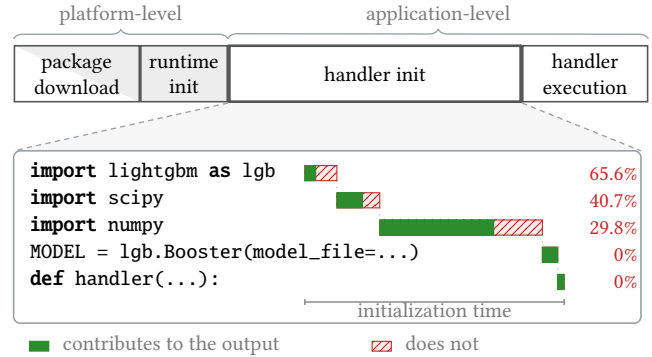
\begin{figure}[t]
  \centering
%
\usetikzlibrary{arrows.meta,calc,patterns,decorations.pathreplacing}%
\definecolor{keepline}{HTML}{2E8B2E}%
\definecolor{passline}{HTML}{9AA0A6}%
\definecolor{steplab}{HTML}{7C7C7C}%
\definecolor{edgecol}{HTML}{555555}%
\definecolor{delline}{HTML}{C62828}%
\definecolor{delfill}{HTML}{FBEAEA}%
\definecolor{provfill}{HTML}{EDEDED}%
\definecolor{funnelfill}{HTML}{F7F7F7}%
\providecommand{\lcfont}{\fontsize{8}{9.6}\selectfont}%
\providecommand{\lckw}[1]{\textbf{#1}}%
\providecommand{\lcseg}[5]{%
  \node[cod] at (0.025\columnwidth,#1) {#2};%
  \fill[delfill]                    (#3\columnwidth,#1-0.055) rectangle (#5\columnwidth,#1+0.145);%
  \pattern[pattern=north east lines, pattern color=delline]
                                    (#3\columnwidth,#1-0.055) rectangle (#5\columnwidth,#1+0.145);%
  \draw[delline, line width=0.4pt]  (#3\columnwidth,#1-0.055) rectangle (#5\columnwidth,#1+0.145);%
  \fill[keepline]                   (#3\columnwidth,#1-0.055) rectangle (#4\columnwidth,#1+0.145);%
  \draw[keepline, line width=0.4pt] (#3\columnwidth,#1-0.055) rectangle (#4\columnwidth,#1+0.145);%
}%
\providecommand{\lcpct}[2]{%
  \node[pct] at (0.975\columnwidth,#1+0.045) {#2};%
}%
\providecommand{\lccarry}[3]{%
  \draw[passline, line width=0.4pt, dotted] (#1\columnwidth,#2) -- (#1\columnwidth,#3);%
}%
\begin{tikzpicture}[
  x=1cm, y=1cm,
  seg/.style={draw=edgecol, line width=0.6pt, inner sep=0pt, anchor=west,
              font=\lcfont, align=center, minimum height=0.74cm},
  prov/.style={seg, fill=provfill},
  dev/.style={seg, fill=white},
  both/.style={seg, fill=white, path picture={
    \fill[provfill] (path picture bounding box.north west)
      -- (path picture bounding box.north east)
      -- (path picture bounding box.south east) -- cycle;}},
  focus/.style={dev, line width=1.0pt},
  lab/.style={font=\lcfont, text=steplab, inner sep=1.5pt},
  pct/.style={font=\lcfont, text=delline, inner sep=0pt, anchor=east},
  cod/.style={font=\lcfont\ttfamily, anchor=base west, inner sep=0pt},
  brc/.style={draw=passline, line width=0.5pt, decorate,
              decoration={brace, amplitude=3.5pt}},
]

\node[both,  minimum width=0.195\columnwidth] (fetch) at (0,0)          {package\\download};
\node[prov,  minimum width=0.135\columnwidth] (rinit) at (fetch.east)   {runtime\\init};
\node[focus, minimum width=0.455\columnwidth] (finit) at (rinit.east)   {handler init};
\node[dev,   minimum width=0.195\columnwidth] (hexec) at (finit.east)   {handler\\execution};

\draw[brc] ($(fetch.north west)+(0,0.08)$) -- ($(rinit.north east)+(0,0.08)$)
  node[midway, above=3.5pt, lab] {platform-level};
\draw[brc] ($(finit.north west)+(0,0.08)$) -- ($(hexec.north east)+(0,0.08)$)
  node[midway, above=3.5pt, lab] {application-level};

\fill[funnelfill]
  (finit.south west) -- (finit.south east) --
  (0.98\columnwidth,-1.02) -- (0,-1.02) -- cycle;
\draw[passline, line width=0.5pt, dashed, dash pattern=on 2pt off 1.6pt]
  (finit.south west) -- (0,-1.02);
\draw[passline, line width=0.5pt, dashed, dash pattern=on 2pt off 1.6pt]
  (finit.south east) -- (0.98\columnwidth,-1.02);

\draw[rounded corners=2.5pt, draw=passline, line width=0.6pt, fill=white]
  (0,-1.02) rectangle (0.98\columnwidth,-3.30);

\lcseg{-1.36}{\lckw{import} lightgbm \lckw{as} lgb}{0.455}{0.472}{0.505}
\lcseg{-1.72}{\lckw{import} scipy}{0.505}{0.545}{0.572}
\lcseg{-2.08}{\lckw{import} numpy}{0.572}{0.750}{0.826}
\lcseg{-2.44}{MODEL = lgb.Booster(model\_file=...)}{0.826}{0.850}{0.850}
\lcseg{-2.80}{\lckw{def} handler(...):}{0.850}{0.860}{0.860}

\lcpct{-1.36}{65.6\%}
\lcpct{-1.72}{40.7\%}
\lcpct{-2.08}{29.8\%}
\lcpct{-2.44}{0\%}
\lcpct{-2.80}{0\%}

\lccarry{0.505}{-1.415}{-1.575}
\lccarry{0.572}{-1.775}{-1.935}
\lccarry{0.826}{-2.135}{-2.295}
\lccarry{0.850}{-2.495}{-2.655}

\draw[draw=steplab, line width=0.5pt] (0.455\columnwidth,-3.02) -- (0.860\columnwidth,-3.02);
\draw[draw=steplab, line width=0.5pt] (0.455\columnwidth,-2.98) -- (0.455\columnwidth,-3.06);
\draw[draw=steplab, line width=0.5pt] (0.860\columnwidth,-2.98) -- (0.860\columnwidth,-3.06);
\node[lab, anchor=base] at (0.6575\columnwidth,-3.22) {initialization time};

\fill[keepline]                   (0.025\columnwidth,-3.67) rectangle (0.055\columnwidth,-3.54);
\draw[keepline, line width=0.4pt] (0.025\columnwidth,-3.67) rectangle (0.055\columnwidth,-3.54);
\node[lab, anchor=base west] at (0.070\columnwidth,-3.65) {contributes to the output};
\fill[delfill]                    (0.560\columnwidth,-3.67) rectangle (0.590\columnwidth,-3.54);
\pattern[pattern=north east lines, pattern color=delline]
                                  (0.560\columnwidth,-3.67) rectangle (0.590\columnwidth,-3.54);
\draw[delline, line width=0.4pt]  (0.560\columnwidth,-3.67) rectangle (0.590\columnwidth,-3.54);
\node[lab, anchor=base west] at (0.605\columnwidth,-3.65) {does not};

\end{tikzpicture}
  \caption{Lifecycle of a serverless function on a cold start.
  A warm invocation starts at \emph{handler execution}.
  The \emph{handler init} phase is expanded
  to show the module-level statements executed
  before the handler is invoked.}
  \label{fig:lifecycle}
\end{figure}

\point{Cold-start latency}
This deployment model introduces
a well-known challenge:
\textit{cold-start latency}.
When no initialized execution environment is available,
the cloud provider must provision a new environment,
fetch the deployment package,
and initialize the runtime,
before invoking the handler.
Providers may keep an initialized environment
for a limited period afterwards,
so that a later request reuses it
and invokes the handler directly
(a~\textit{warm start}).
Figure~\ref{fig:lifecycle} shows
the cold-start lifecycle of a real-world Python handler.
Its initialization phase
loads a {\tt lightgbm} model,
executing the top-level statements
of the handler's module
and of every module its imports transitively
load for the first time.

Cold-start latency is among the primary concerns
for cloud developers and operators~\cite{datadog-2023}.
First,
a cold start delays the response observed by the user,
increasing latency by up to 80\%
over a warm invocation~\cite{l-trim}.
Since most serverless applications
have latency requirements~\cite{eismann2021serverless},
this delay matters
even when cold starts affect
only a fraction of a handler's requests:
they worsen tail latency,
which is often used to define
service-level objectives~\cite{slimstart,joosen2025eurosys}.
Second,
initialization can also increase
execution costs.
When initialization time is billed,
developers pay for both the
initialization phase
(Figure~\ref{fig:lifecycle})
and the handler's execution~\cite{aws-lambda-pricing,aws-init-billing}.
Initialization accounts for 53.8\%
of the billed duration
for the median Python handler
in~\cite{l-trim}.

\point{Contributing factors}
Several factors contribute to cold-start latency.
\textit{Platform-level} factors include
the time the provider spends provisioning
execution environments,
scheduling them,
and allocating resources.
\textit{Application-level} factors
are those developers can influence,
namely the deployment package
and the work done during handler initialization.
Both can contain unnecessary components.
Applications often depend on large dependency trees,
and more than 95\% of the library functions they ship
are never used~\cite{bloat-study}.
However,
unused code is only part of the problem.
Prior work identifies code that \emph{does} execute
during initialization
and~\emph{never} influences the handler's behavior,
such as third-party setup
that runs as a side effect of importing a library~\cite{l-trim,slimstart}.
In Figure~\ref{fig:lifecycle},
this waste accounts for 30\% to 66\%
of the initialization each import triggers.
Providers accordingly advise developers to trim
dependencies and initialization
work~\cite{aws-python-best-practices,gcp-functions-best-practices}.

\point{Limitations}
Existing application-level approaches
have several limitations.
\ltrim~\cite{l-trim} debloats an application
by delta debugging~\cite{delta-debugging}
over the top-level statements of one module at a time,
so it cannot remove code whose deletion
requires coordinated changes across modules.
FaaSLight~\cite{faaslight} stubs the bodies
of statically unreachable functions
and loads them on demand,
to avoid compiling code that never runs.
This is a wrong assumption,
since deployed applications ship with pre-compiled bytecode.
Worse,
a stub called during initialization
must compile its body from source,
making cold starts slower rather than faster.
\slimstart~\cite{slimstart} defers expensive imports
until their first use,
which delays their side effects
and can change observable
behavior~\cite{cinder-lazy-imports}.

\point{Approach}
We instead view debloating
as a dynamic program slicing~\cite{weiser84slicing,korel88dynamic,agrawal90dynamic}
problem,
where the slicing criterion is the
set of operations within a handler
that produce externally visible effects.
We realize this in~\xtrim,
which computes the slice
by observing the handler's execution
on a given set of workloads.
A~\textit{shadow interpreter} runs alongside CPython,
recording the data and control dependences
of every executed instruction.
Because native extensions are invisible to it,
a~\textit{C-API interceptor} patches
the dispatch tables of loaded extensions,
recording every Python object
that native code reads or writes.
The same interception identifies the operations
that reach the outside world
and form the criterion.
\xtrim\ then deletes every statement
on which no such operation depends,
directly or transitively,
leaving an application that behaves identically
on the observed workloads
but loads and executes far less code.

\point{Results}
On~\numberofapps\ serverless applications deployed on AWS Lambda,
\xtrim\ reduces cold-start latency
by \empirical{21.7\%}
and peak memory by~\empirical{17.1\%} at the median,
and never makes either worse.
This is more than twice the reduction of~\ltrim,
which achieves~\empirical{8.5\%}
and~\empirical{5.4\%} reductions
on the same applications.
\xtrim\ is also twelve times faster at
debloating
(\empirical{5} minutes versus~\empirical{62} at the median),
because~\ltrim\ re-executes the application
for every candidate removal
while~\xtrim\ needs a single profiling run.
An ablation confirms that both parts
of the analysis are needed.
Replacing our criterion with a proxy
that keeps whatever is read
halves the reduction,
and disabling the C-API interceptor
breaks 19 of the 31 applications
because of uncaptured dependences.

\point{Contributions}
We make the following contributions:
\begin{itemize}
\item
Conceptually,
we formulate cold-start minimization
as a dynamic slicing problem,
with a criterion that captures
a handler's externally visible behavior
instead of approximating it by static reachability
or output equivalence.
(Section~\ref{sec:approach}).

\item
Technically,
we present~\xtrim,
a realization of the approach for Python.
It records dependences over
CPython bytecode and recovers those
created inside native extensions,
so it handles real-world handlers
that use dynamic features
and interoperate with compiled code.
(Section~\ref{sec:system}).

\item
Empirically,
we evaluate~\xtrim\ on~\numberofapps\ applications,
running~\ltrim,
FaaSLight,
and~\slimstart\ on the same suite.
Against the original handlers,
\xtrim\ reduces cold-start latency
by~\empirical{21.7\%}
and peak memory by~\empirical{17.1\%} at the median,
more than twice~\ltrim's reduction
(Section~\ref{sec:evaluation}).
\end{itemize}

\section{Background and Motivation}
\label{sec:background}

We state the problem of cold-start minimization,
and list its challenges.
We then discuss the limitations of prior work.


\newlength{\lcbandw}
\settowidth{\lcbandw}{\footnotesize\ttfamily xxxxxxxxxxxxxxxxxxxxxxxxxxxxxxxxxxxxx}
\begin{figure*}[t]
  \centering
  \begin{subfigure}[t]{0.36\textwidth}
\begin{lstlisting}[language=Python, numbers=none, xleftmargin=4pt,
    basicstyle=\footnotesize\ttfamily, keepspaces=true, aboveskip=2pt, belowskip=2pt,
    escapeinside={(*@}{@*)}]
# main.py
import plugins                   # m1
import registry                  # m2
import legacy                    # m3

def handler(name):               # m4
    fn = registry.HOOKS["title"] # m5
    n = fn(name)                 # m6
    parser = legacy.parse        # m7
    return n                     # m8

(*@{\setlength{\fboxsep}{1.5pt}\hspace*{-\fboxsep}\colorbox{black!9}{\makebox[\lcbandw][l]{handler(\textcolor[rgb]{0,0,1}{"ada"})\hfill\textcolor[rgb]{0.133,0.545,0.133}{\# i1}}}}@*)
\end{lstlisting}
    \caption{}\label{fig:run-main}
  \end{subfigure}
  \hfill
  \begin{subfigure}[t]{0.36\textwidth}
\begin{lstlisting}[language=Python, numbers=none, xleftmargin=4pt,
    basicstyle=\footnotesize\ttfamily, keepspaces=true, aboveskip=2pt, belowskip=2pt]
# registry.py
HOOKS = {}                       # r1

# plugins.py
import registry                  # p1
def title(s):                    # p2
    return s.title()             # p3
registry.HOOKS["title"] = title  # p4

# legacy.py
import xmlkit                    # l1
def parse(doc):                  # l2
    return xmlkit.parse(doc)     # l3
\end{lstlisting}
    \caption{}\label{fig:run-modules}
  \end{subfigure}
  \hfill
  \begin{subfigure}[t]{0.26\textwidth}
    \raisebox{-\height}{\includegraphics[width=0.93\linewidth]{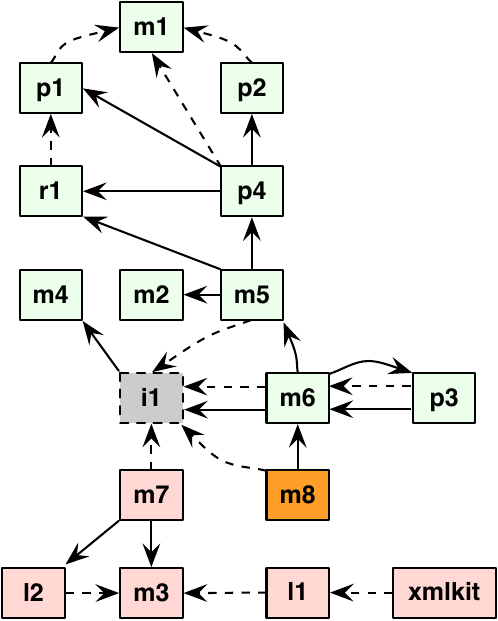}}
    \caption{}\label{fig:ddg}
  \end{subfigure}
  \caption{
	(a) and (b) show a Python handler and its dependencies.
	The highlighted line {\tt i1} is the platform
	invoking the handler on a workload.
	(c) shows the resulting dynamic dependence graph
	discussed in Section~\ref{sec:approach}:
	solid edges denote data dependences, and
	dashed edges denote control dependences.
	The orange node is the sink,
	red nodes correspond to unneeded steps.}
  \label{fig:running-ex}
\end{figure*}

\point{Problem statement}
Our goal is
to reduce cold-start latency by
focusing on the~\textit{application-level} factors:
\textit{
identifying and removing statements
in the handler's initialization code
and its dependency tree
that either
(1) never execute,
or (2) execute but never influence
the handler's observable behavior.
}
We call this process~\textit{debloating}.
We focus on removing import statements
and function definitions,
because initialization code consists
almost entirely of these.
Imports are the most impactful target,
since each one triggers the execution of another module's top-level code,
so removing a single redundant import can eliminate
hundreds or thousands of executed statements.
Overall,
debloating brings two benefits:
(1) the deployment package becomes smaller,
reducing download time,
and (2) less code executes during initialization,
reducing handler initialization time.

\point{Running Example}
Figure~\ref{fig:running-ex}
shows the running example
we use throughout the paper
to explain our approach
(let us ignore the graph for now).
It is a Python handler
whose top-level code imports
a couple of dependencies,
not all of which are needed for its
observable behavior.
The module {\tt legacy}
is imported at {\tt m3}
and its function {\tt legacy.parse} is referenced
and bound to the variable {\tt parser} at {\tt m7},
but it is never actually called.
Consequently,
statement {\tt l3} is never executed,
and while statements {\tt m3},
{\tt m7},
and {\tt l1} do execute,
they have no impact on the handler’s output.

\point{Challenges}
Automatically
identifying and removing such redundant statements
(i.e., {\tt import legacy})
involves several challenges.

\textit{\textbf{C1} Dependent statements.}
Removing a statement is not a local decision.
Every statement that transitively depends on
a removed one must be removed too.
In Figure~\ref{fig:run-main},
deleting the import at {\tt m3} requires
deleting {\tt m7}.

\textit{\textbf{C2} Side effects and ordering.}
A statement can be needed without
producing any value the handler consumes.
The module {\tt plugins} is imported at {\tt m1}
(Figure~\ref{fig:run-main})
and never referenced again in {\tt main.py}.
However,
its top-level code registers
{\tt title} in {\tt registry.HOOKS} at {\tt p4}
(Figure~\ref{fig:run-modules}),
which the handler reads at {\tt m5}.
Removing the import would break the handler,
although no statement uses the name {\tt plugins}.
Ordering matters too.
Deferring the import elsewhere
can make the write at {\tt p4} occur
after the read at {\tt m5},
so the handler fails with {\tt KeyError}.

\textit{\textbf{C3} Native extensions.}
Python handlers
routinely depend on other packages
(e.g., {\tt numpy}, {\tt torch})
containing code written in C or Rust~\cite{pyxray}.
These native extensions
can read and write Python objects
and import other Python modules,
without any of these operations
appearing in the Python code.
A debloater must consider both sides
of the language boundary,
or it misses these dependences
and removes needed code.

\textit{\textbf{C4} Handler's observable behaviors.}
Before deciding what to remove,
we must know what the handler is expected to produce.
The handler's return value is one part,
but a handler may also write to storage,
call a remote service
or log.
Unlike the return value,
operations that influence
the handler's observable behavior
have no syntactic marker
and are scattered through the application
and its dependencies.

\point{Existing work and limitations}
Existing work leaves these challenges unaddressed.
\ltrim~\cite{l-trim} has no notion of dependences
(\textbf{C1}).
Removing {\tt m3} leaves {\tt m7} untouched,
the handler crashes with a {\tt NameError},
and~\ltrim\ reverts the removal,
keeping the unneeded import of {\tt legacy}.
\slimstart~\cite{slimstart} defers an import
together with the side effects
of the module it loads
(\textbf{C2}).
Making the import of {\tt plugins} lazy means
{\tt p4} never registers {\tt title},
and the handler fails with a {\tt KeyError} at {\tt m5}.
Such side effects are common,
since widely used Python packages rely on
them heavily~\cite{cinder-lazy-imports}
and they are hard to detect.
FaaSLight~\cite{faaslight} avoids a compilation cost
that deployed applications never pay,
so it brings no benefit and makes cold starts worse
(Section~\ref{sec:evaluation}).

\section{Debloating as a Dynamic Slicing Problem}
\label{sec:approach}

Motivated by these challenges
and the limitations of existing work,
we now formulate our debloating approach.
Serverless applications
are written predominantly in
dynamic languages~\cite{datadog-2023,joosen2025eurosys},
where reflection
makes a static approach unable to prove
either that code is unreachable
or that reachable code is unneeded.
Therefore,
we take a dynamic approach,
observing the handler's execution.
Our insight is that
debloating can be cast as
\textit{dynamic slicing}~\cite{agrawal90dynamic,hrb1990},
a program analysis technique
that computes the statements directly
or transitively affecting a given~\textit{criterion}.

\point{Dynamic dependence graph}
\label{sec:ddg}
Our formulation relies on
the notion of a~\textit{dynamic dependence graph (DDG)}~\cite{agrawal90dynamic},
which captures the dependences (defined below)
among a program's statement instances.
A statement instance,
\textit{step} for short,
is one execution occurrence of a statement:
the statement is syntax,
written once,
while a step is one of the times it actually ran.
For example,
a statement in a loop body contributes one step per iteration.

Formally,
a DDG is defined as
$\mathit{DDG} = (V, E)$.
The nodes $V$ are the steps of the execution,
and the edges $E \subseteq V \times V \times L$
are labeled with a dependence kind
taken from $L = \{\texttt{d}, \texttt{c}\}$.
We write $s \overset{l}{\rightarrow} s'$
for an edge $(s, s', l) \in E$,
and the two labels capture
the following relationships between steps.

\textit{Data dependences.}
A step $s$ is data-dependent on step $s'$
(denoted as $s \overset{\texttt{d}}{\rightarrow} s'$),
if $s$ reads a cell
(e.g., a variable, a memory location)
and $s'$ is the most recent step
preceding $s$ in the execution
that writes to that cell.

\textit{Control dependences.}
A step $s$ is control-dependent on step $s'$
(denoted as $s \overset{\texttt{c}}{\rightarrow} s'$),
if $s'$ decides whether $s$ executes.
This arises in two cases:
$s'$ is the test of the nearest enclosing branch or loop
that lets $s$ execute,
or $s$ belongs to the body of a function
and $s'$ is the call step that invoked it.

\point{Example}
Figure~\ref{fig:ddg} shows the DDG
for our running example.
Step \texttt{m8} is control-dependent on \texttt{i1},
the call the platform issues,
while \texttt{m6} is data-dependent on it,
since the argument \texttt{name} it reads
is bound by that call.
Our definition treats an import as a call,
since it triggers the execution
of a module's top-level statements.
The graph contains only steps that actually occurred,
which is why {\tt l3} is absent.

\point{Slicing formulation}
\label{sec:dynamic-slicing}
A DDG turns the question of
which statements are needed into a reachability query over its steps.
The query needs a~\textit{criterion},
a set of steps that must be present in
the debloated program no matter what.
This translates to the steps
whose effects are visible outside the handler,
since removing one of those would change
what the caller of the handler observes.
We call these steps~\textit{sinks}
and assume for now that they are given
(Section~\ref{sec:sinks}).
Given a set of sinks $T$,
the~\textit{backward slice}
($\mathit{bslice}(T)$)
is the set of steps the sinks depend on,
directly or transitively.
This corresponds to the minimal set
that must be kept.

However,
not every statement contributes
equally to cold-start latency,
which is dominated by
the~\textit{initialization
of unneeded dependency code}~\cite{l-trim,slimstart,faaslight,pep8102025,cinder-lazy-imports,aws-init-billing}.
That cost is paid by import statements
and by the module top-level code they execute,
most of which consists of function definitions.
We therefore restrict removal
to these two kinds of statements.
We call their steps~\textit{sources} $S$
and compute the forward slice
$\mathit{fslice}(S)$,
i.e.,
the steps that depend on a source,
directly or transitively.
Since anything we remove lies in $\mathit{fslice}(S)$,
we trace forward from $S$
and build the subgraph it induces.

\point{Debloating}
The two slices yield the following formulation.
\begin{definition}[Debloating]
\label{def:debloating}
Given a program $P$,
a set of workloads $W$,
a set of sources $S$,
and a set of sinks $T$,
the \emph{unneeded statements} $R \subseteq P$
are the largest set such that,
for every $\sigma \in R$:
\begin{enumerate}[label=(\arabic*),leftmargin=*,nosep]
\item $\mathit{steps}(\sigma) \subseteq
       \mathit{fslice}(S) \setminus \mathit{bslice}(T)$; and
\item $\mathit{stmt}(s') \in R$
      for every edge $s' \overset{l}{\rightarrow} s$
      with $s \in \mathit{steps}(\sigma)$,
\end{enumerate}
where $\mathit{steps}(\sigma)$ denotes
the steps of statement $\sigma$ across these executions
and $\mathit{stmt}(s)$ the statement of step $s$.

The debloated program is then given by
$P' = P \setminus R$,
and the \emph{unneeded steps} $U$
are the steps of the statements in $R$.
\end{definition}
Condition (1) makes a statement a candidate
when it originates at a source
and contributes nothing to a sink,
and condition (2) removes a candidate
only if every statement depending on it
is unneeded too.
Through these conditions,
the debloated program is an executable subprogram
that reproduces $P$'s computations
on each workload in $W$~\cite{korel88dynamic}.
Statements that~\emph{never} execute
have no steps and no dependents,
so they are removed too.
In Figure~\ref{fig:ddg},
$U$ is the set of red nodes,
so {\tt import legacy} is removed.

\point{Guarantees and assumptions}
Our debloating formulation addresses
challenges~\textbf{C1} and~\textbf{C2}
(Section~\ref{sec:background})
by construction.
No surviving statement can depend on
a removed one (\textbf{C1}),
since a statement's dependents lie in the same forward slice,
and condition (2) requires them to be removed together with it.
Side effects
(\textbf{C2})
need no special treatment,
as a heap update is a write to a cell,
so its later uses appear
as data dependences.
Ordering is also preserved because
deletion does not reorder the statements
in the debloated program.
For example,
in Figure~\ref{fig:ddg},
the sink
({\tt m8})
needs the hook that \texttt{plugins.py}
stored in \texttt{registry.HOOKS} at import time
({\tt p4}),
captured by the path
\texttt{m5}~$\overset{\texttt{d}}{\rightarrow}$~\texttt{p4}~$\overset{\texttt{c}}{\rightarrow}$~\texttt{m1},
so the import at \texttt{m1} is kept,
even though \texttt{main.py}
never calls \texttt{plugins} directly.

Challenges~\textbf{C3} and~\textbf{C4}
concern realizing the model on an actual runtime,
which Section~\ref{sec:system} addresses.

\section{System Design}
\label{sec:system}

\begin{figure}[t]
  \begin{center}
    \includegraphics[width=\linewidth]{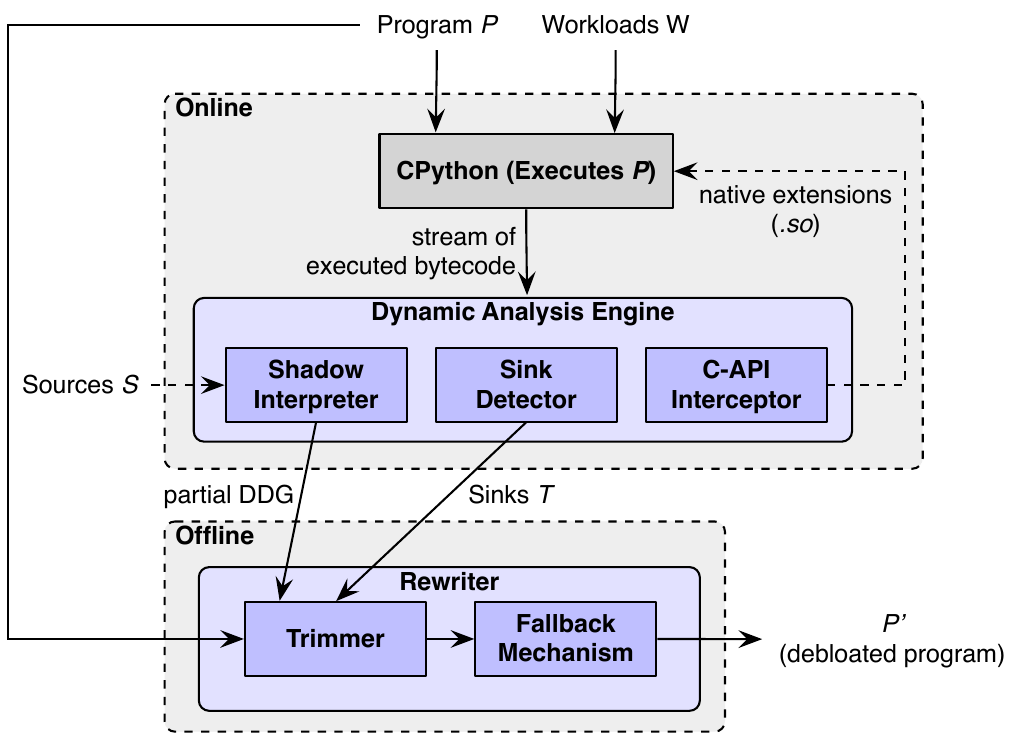}
  \end{center}
  \caption{High-level architecture of~\xtrim.}
  \label{fig:arch}
\end{figure}

\newlength{\lcbandwb}
\settowidth{\lcbandwb}{\footnotesize\ttfamily xxxxxxxxxxxxxxxxxxxxxxxxxxxx}

\begin{figure*}[t]
  \centering
  \begin{minipage}[b]{0.32\textwidth}
    \begin{subfigure}[b]{\linewidth}
\begin{lstlisting}[language=PythonBytecode, xleftmargin=17pt,
    basicstyle=\footnotesize\ttfamily, keepspaces=true,
    escapeinside={(*@}{@*)}]
     RESUME
p1:  LOAD_CONST 0
     LOAD_CONST None
     IMPORT_NAME registry
     STORE_NAME registry
p2:  LOAD_CONST <code title>
     MAKE_FUNCTION
     STORE_NAME title
p4:  LOAD_NAME title
     LOAD_NAME registry
     LOAD_ATTR HOOKS
     LOAD_CONST 'title'
     STORE_SUBSCR
(*@{\setlength{\fboxsep}{1.5pt}\hspace*{-\fboxsep}\colorbox{black!9}{\makebox[\lcbandwb][l]{\hphantom{xxxxx}\textbf{RETURN\_CONST} None}}}@*)
\end{lstlisting}
      \caption{Bytecode of \texttt{plugins}.}
      \label{fig:state-bytecode}
    \end{subfigure}
  \end{minipage}%
  \hfill
  \begin{minipage}[b]{0.66\textwidth}
    \begin{subfigure}[b]{\linewidth}
      \centering
      \includegraphics[width=0.83\linewidth]{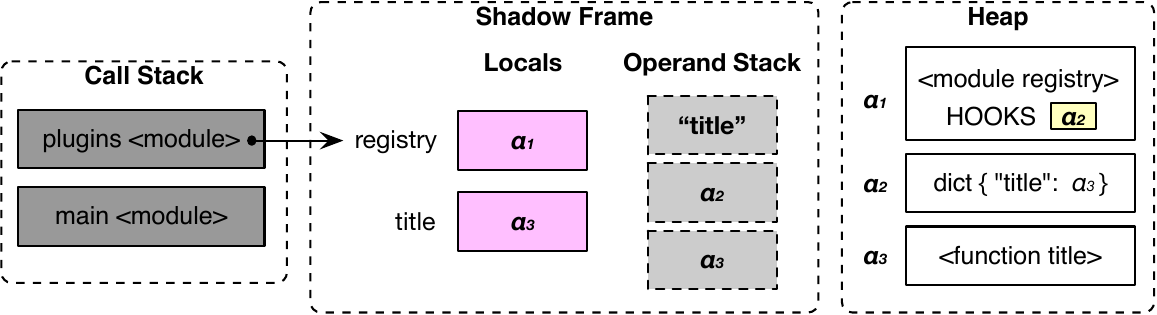}
      \caption{CPython interpreter state.}
      \label{fig:state-plain}
    \end{subfigure}

    \medskip
    \begin{subfigure}[b]{0.57\linewidth}
      \centering
      \includegraphics[width=0.95\linewidth]{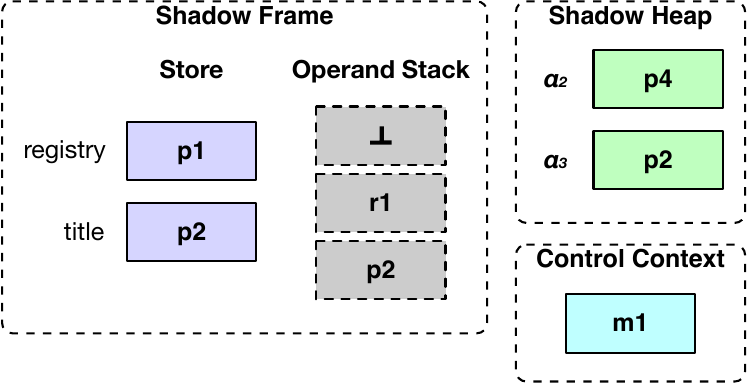}
      \caption{Shadow state maintained by \xtrim.}
      \label{fig:state-shadow}
    \end{subfigure}%
    \hfill
    \begin{subfigure}[b]{0.40\linewidth}
      \centering
      \includegraphics[width=0.95\linewidth]{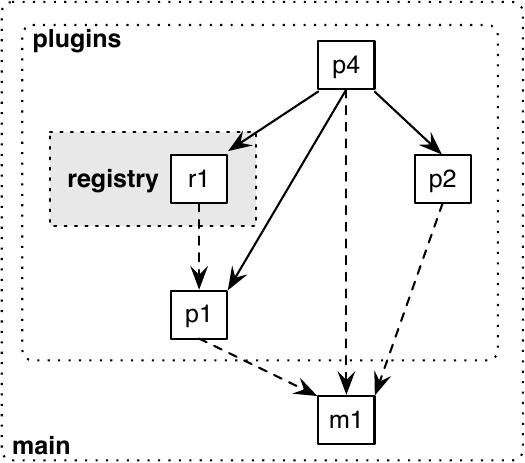}
      \caption{Partial DDG recorded so far.}
      \label{fig:state-graph}
    \end{subfigure}
  \end{minipage}
  \caption{Interpreter and shadow state
  while performing the import at \texttt{m1},
  which loads module \texttt{plugins}
  (Figure~\ref{fig:run-modules}).
  The state is captured at step \texttt{p4},
  right after \texttt{STORE\_SUBSCR} updates {\tt registry.HOOKS}.
  The greyed slots are the operands it popped.
  In (d),
  shaded nodes come from the imported module {\tt registry}
  ({\tt r1}, Figure~\ref{fig:running-ex}).}
  \label{fig:state}
\end{figure*}

\point{Overview}
\xtrim\ realizes the debloating process
of Section~\ref{sec:approach},
and is shown in Figure~\ref{fig:arch}.
It takes a Python serverless application $P$,
together with its dependencies,
and a set of workloads $W$,
and proceeds in two phases.
In the~\textit{online} phase,
\xtrim\ runs $P$ on each workload in $W$
under the stock CPython interpreter,
while a~\textit{dynamic analysis engine}
monitors the execution at the level of bytecode.
The engine has three components.
A~\textit{shadow interpreter}
(Section~\ref{sec:shadow-interpreter})
processes every executed instruction
and computes the partial DDG
induced by $\mathit{fslice}(S)$ of the given sources $S$.
A~\textit{sink detector}
(Section~\ref{sec:sinks})
identifies the sinks on the fly,
since, unlike the sources,
they are not known syntactically (\textbf{C4}).
A~\textit{C-API interceptor}
(Section~\ref{sec:c-api-interceptor})
records the heap accesses that native code performs (\textbf{C3}),
which the shadow interpreter turns into
data and control dependences.

In the~\textit{offline} phase,
a~\textit{rewriter} deletes the unneeded statements
from the application and its dependencies.
Its~\textit{trimmer}
(Section~\ref{sec:trimmer})
traverses the recorded DDG backwards from the sinks
to obtain $\mathit{bslice}(T)$
and applies Definition~\ref{def:debloating},
while its~\textit{fallback mechanism}
re-invokes the original application
for inputs that reach removed code.
The result is a debloated program $P'$,
which is deployed to the cloud.


\subsection{Shadow Interpreter}
\label{sec:shadow-interpreter}

To build the partial DDG,
the shadow interpreter propagates
\textit{labels} over Python bytecode.
We work at this level
because a single statement may perform
several independent reads and writes,
and line-level tracing cannot tell which of them the result depends on.
Bytecode makes each read and write a separate instruction,
so each dependence is attributed to the operation
that created it.
Figure~\ref{fig:state-bytecode} shows
the bytecode compiled from the {\tt plugins.py} module of
Figure~\ref{fig:run-modules},
which we use as an example throughout this section.

\point{Labels}
To build the partial DDG,
the shadow interpreter must know 
which step produced a value 
whenever an instruction consumes it.
It records this by attaching to each value a label,
which is the step that produced it.
A step is a source location~\cite{pep-source-loc}
together with an occurrence number,
since one location may execute many times.

\point{State}
The shadow interpreter mirrors
CPython's state.
Wherever the real interpreter holds a value,
the shadow holds that value's label.
A~\textit{shadow frame},
one per real frame
(a function invocation
or a module's top-level execution),
contains
(1) a shadow store
mapping each local name to the label of its current value,
and (2) a shadow operand stack
holding one label per real operand stack slot.
A~\textit{shadow heap} mimics CPython's heap,
mapping addresses of objects that outlive a frame
(e.g., object attributes)
to the label of the step that last wrote them.
A control context records
why the current instruction is executing.
It is a sequence of steps
in which each step caused the next to be reached,
including the calls and imports that entered
the frames on the call stack,
interleaved with the predicates of the branches in effect.
Figure~\ref{fig:state-shadow} shows
the shadow interpreter's state
just after the {\tt STORE\_SUBSCR} at line 13
(Figure~\ref{fig:state-bytecode}),
which performs the update of the {\tt registry.HOOKS} dictionary
({\tt p4}, Figure~\ref{fig:run-modules}).
According to the shadow store,
the local variable {\tt registry} carries
the label $\texttt{p1}$,
recording that its value was produced
by the import statement at {\tt p1}
(Figure~\ref{fig:run-modules}).

\point{Sources}
Labels are not created everywhere.
Since our goal is to find the statements
that depend on imports and function definitions
(Section~\ref{sec:background}),
only two opcodes introduce them.
Every form of Python import
(e.g., {\tt import m as n}) compiles to
an {\tt IMPORT\_NAME}
(line 4, Figure~\ref{fig:state-bytecode}),
and every function definition and lambda to
a {\tt MAKE\_FUNCTION}
(line 7, Figure~\ref{fig:state-bytecode}).
Executing either labels the value it produces
with the executing step.
We call these instructions sources.
No other instruction introduces a label,
so a value carries one only if it descends
from an import or a function definition.

\point{Recording dependences}
The shadow interpreter augments
CPython's operational semantics
with rules that operate on labels.
Whenever an instruction at step $s$ reads
a label $\ell$ from the shadow operand stack,
the shadow store,
or the shadow heap,
it records a data dependence
from $s$ to $\ell$.

Regarding control dependences,
the shadow interpreter records them lazily,
acting only at stores that are~\textit{escaping},
meaning they write to a cell that outlives the current frame.
Examples include
(1) a store to a name in a module's namespace,
such as {\tt registry} at {\tt p1}
(line 5,
Figure~\ref{fig:state-bytecode}),
because a module's namespace stays reachable via
{\tt sys.modules},
or (2) a store to an attribute of a heap object,
such as {\tt registry.HOOKS} at {\tt p4}
(line 13,
Figure~\ref{fig:state-bytecode}).

To record why an escaping store executed,
the interpreter reads the control context,
which holds the steps that led to $s$.
An escaping store at $s$
is control-dependent on
the innermost step of the control context,
that step on the one enclosing it,
and so on outward.
Writing the control context as
$\langle \ell_1,\dots,\ell_k\rangle$
from outermost to innermost,
the engine adds the edges
$s \overset{\texttt{c}}{\rightarrow} \ell_k$
and $\ell_{i + 1} \overset{\texttt{c}}{\rightarrow} \ell_i$
for every $i < k$,
a set of edges
denoted as $\mathit{chain}(s)$.
Because the control context spans the whole call stack,
$\textit{chain}(s)$ makes $s$ reach the import or call
that triggered the escaping store.
Appendix~\ref{appendix:rules}
gives representative rules.

\point{Complete example}
Figure~\ref{fig:state} traces
the top level of {\tt plugins}
up to the {\tt STORE\_SUBSCR} at line 13.
The sources at lines 4 and 7 create the labels
{\tt p1} and {\tt p2},
which the following {\tt STORE\_NAME} instructions
bind to {\tt registry} and {\tt title} in the shadow store.
Both stores are module-level,
so their cells escape.
Since there is no enclosing branch,
each receives a single control edge to {\tt m1}
(Figure~\ref{fig:running-ex}),
which corresponds to the
statement that entered the module.

Step {\tt p4} then updates {\tt registry.HOOKS}.
The {\tt LOAD\_ATTR} at line 11 reads {\tt registry},
records a data edge to its label {\tt p1},
and pushes {\tt r1},
the label the shadow heap holds for {\tt registry.HOOKS}.
The {\tt STORE\_SUBSCR} at line 13 consumes
{\tt r1} and {\tt p2}
($\bot$ for the constant key, which carries no label),
records a data edge to each,
and relabels that cell to {\tt p4}.
The cell belongs to a heap object,
so this store escapes too,
and {\tt p4} receives a control edge to {\tt m1} as well.

\subsection{C-API Interceptor}
\label{sec:c-api-interceptor}

\begin{figure}[t]
  \begin{center}
    \includegraphics[width=\linewidth]{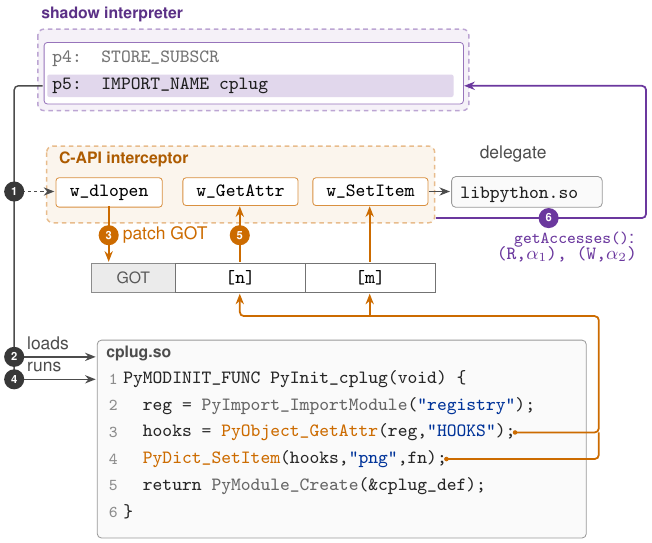}
  \end{center}
  \caption{
      High-level overview of our C-API interceptor
      when importing a native extension called {\tt cplug.so}.
	  The example assumes that {\tt cplug} is imported
	  after line {\tt p4} of Figure~\ref{fig:run-modules}.
  }
  \label{fig:interceptor}
\end{figure}

\point{The problem}
Our shadow interpreter
(Section~\ref{sec:shadow-interpreter})
observes only Python bytecode,
yet serverless applications
routinely employ native extensions
(\textbf{C3}, Section~\ref{sec:background})
whose code is opaque to it.
Native code is entered in two ways:
(1) a~\textit{native call},
a {\tt CALL} instruction whose callee has no bytecode of its own,
and (2) a~\textit{native import},
an {\tt IMPORT\_NAME} that resolves to
a compiled module ({\tt .so}) rather than a {\tt .py} file.
For the latter,
CPython loads the shared library with {\tt dlopen}
and runs its initialization function {\tt PyInit\_<name>},
which does what a Python module does
with its top-level statements.
In both cases,
a natural first attempt is to over-approximate:
record a data dependence from the entry point
to every labeled argument it receives.
This is not sufficient,
because an extension can reach back into the Python heap
by importing modules on its own
and accessing objects within them.
Those accesses are invisible to the shadow interpreter,
so the dependences they create are missed,
leading to the removal of statements that are needed
(Section~\ref{sec:trimmer}).

\point{Interception}
What an extension can touch is not arbitrary:
every attribute it reads,
every module it imports,
every object it accesses
must go through CPython's C API~\cite{c-api}
(e.g., {\tt PyObject\_GetAttr},
{\tt PyImport\_Import}),
and those calls are dispatched
through the extension's~\textit{global offset table} (GOT).
Our interceptor exploits this by rewriting the GOT,
as Figure~\ref{fig:interceptor} shows
for an extension {\tt cplug.so}
imported by the module {\tt plugins}.
A GOT exists only once its extension is loaded,
so our engine wraps {\tt dlopen} at startup~(\circled{1}).
The import at {\tt p5} therefore enters
our {\tt w\_dlopen},
which loads the library~(\circled{2})
and patches its GOT~(\circled{3}),
replacing the entries of the API functions
that access Python objects with our own wrappers.
When CPython then runs {\tt PyInit\_cplug}~(\circled{4}),
its API calls resolve to those wrappers~(\circled{5}),
each of which records the access
and delegates to the original function.
The shadow interpreter obtains the recorded accesses
through {\tt getAccesses()}~(\circled{6}):
here the set $\{(R,\alpha_1),(W,\alpha_2)\}$
contains the addresses of
(1) the module {\tt registry} the extension read
and (2) the dictionary {\tt registry.HOOKS} it wrote.

\point{Modeling native code}
The shadow interpreter treats both entry points uniformly,
as a single opaque step $s$
that depends on everything it read
and produces everything it wrote.
Let $R$ and $W$ be the addresses
read and written during the native execution.
For every address in $R \cup W$
whose shadow heap cell holds a label,
the interpreter adds a data dependence from $s$ to it.
For every address in $W$,
it writes $s$ into the cell.
Finally,
it adds the edges in $\textit{chain}(s)$,
exactly as for an escaping store
(Section~\ref{sec:shadow-interpreter}).
In Figure~\ref{fig:interceptor},
only $\alpha_2$ holds a label,
namely {\tt p4},
so {\tt p5} gains a data dependence on {\tt p4},
the cell at $\alpha_2$ is relabeled to {\tt p5},
and $\textit{chain}(\texttt{p5})$ adds
a control dependence to {\tt m1}.


\point{Builtin functions}
CPython's built-in functions
and the methods of built-in types
(e.g., {\tt len},
{\tt list.append})
are compiled into the interpreter
and do not call through the C API,
so the interceptor has nothing to patch.
For these,
the shadow interpreter applies transfer rules
derived from the Python documentation.
For example,
{\tt list.append(lst, x)}
relabels the cell of {\tt lst}
and adds $\textit{chain}(s)$ if the write escapes.
Appendix~\ref{appendix:transfer-rules}
lists representative rules.

\subsection{Sink Detector}
\label{sec:sinks}

The trimmer
(Section~\ref{sec:trimmer})
also needs the sinks,
that is,
the steps whose effects are observable
outside the handler.
Effects escape in two ways.
First,
through the handler's~\textit{return value}
or an uncaught exception,
which the runtime reports to the caller.
Such sinks are known statically:
the {\tt RETURN\_VALUE} and {\tt RETURN\_CONST} instructions
of the handler's body,
and the instructions that propagate an uncaught exception.
Second,
through~\textit{interaction with the outside world},
such as a call to a remote service
or a write to storage.
Such sinks are not known statically,
but our key insight is that
\textit{a Python program can only reach
the outside world by leaving Python}:
every file operation,
socket operation,
and write to stdout is ultimately performed by {\tt libc}.

That boundary is the one
the C-API interceptor
(Section~\ref{sec:c-api-interceptor})
already patches,
so we extend it,
wrapping the {\tt libc} functions
for the file system
({\tt open}, {\tt write})
and for sockets
({\tt socket}, {\tt send}, {\tt connect})
in the GOT of {\tt libpython.so}
as well as in that of every loaded extension.
Its output then reports
whether a native call produced an external effect,
and the shadow interpreter marks that step as a sink.

\subsection{Trimmer and Fallback Mechanism}
\label{sec:trimmer}

After the online phase
(Figure~\ref{fig:arch}),
\xtrim\ proceeds to trimming.
This phase runs offline on the resulting partial DDG.
\xtrim\ parses each Python source file into
an abstract syntax tree (AST)
and deletes the unneeded statements
of Definition~\ref{def:debloating}.
It computes the unneeded statements $R$
as a greatest fixpoint.
Starting from the statements whose steps
all lie in $\mathit{fslice}(S) \setminus \mathit{bslice}(T)$,
it iteratively drops every statement
violating condition (2)
(Definition~\ref{def:debloating})
until it converges.

Deletion is sound for everything the DDG carries,
but two cases in Python's semantics
fall outside Definition~\ref{def:debloating}.

\point{Constructs with compilation-time effects}
Some Python constructs influence
how their enclosing function is compiled,
regardless of whether execution reaches them.
The fragments below illustrate this with {\tt global}.

\begin{center}
\begin{minipage}[c]{0.42\linewidth}%
\begin{lstlisting}[language=Python, numbers=none, xleftmargin=0pt,
    basicstyle=\footnotesize\ttfamily, columns=fixed, keepspaces=true,
    aboveskip=0pt, belowskip=0pt]
def foo():     # s1
  global avar  # s2
  avar = 1     # s3
\end{lstlisting}%
\end{minipage}%
\;$\longrightarrow$\;
\begin{minipage}[c]{0.44\linewidth}%
\begin{lstlisting}[language=PythonBytecode, numbers=none, xleftmargin=0pt,
    basicstyle=\footnotesize\ttfamily, columns=fixed, keepspaces=true,
    aboveskip=0pt, belowskip=0pt]
s1:  RESUME
s3:  LOAD_CONST 1
     STORE_GLOBAL avar
\end{lstlisting}%
\end{minipage}\\[5pt]
\begin{minipage}[c]{0.42\linewidth}%
\begin{lstlisting}[language=Python, numbers=none, xleftmargin=0pt,
    basicstyle=\footnotesize\ttfamily, columns=fixed, keepspaces=true,
    aboveskip=0pt, belowskip=0pt]
def foo():     # s1
    avar = 1   # s2
\end{lstlisting}%
\end{minipage}%
\;$\longrightarrow$\;
\begin{minipage}[c]{0.44\linewidth}%
\begin{lstlisting}[language=PythonBytecode, numbers=none, xleftmargin=0pt,
    basicstyle=\footnotesize\ttfamily, columns=fixed, keepspaces=true,
    aboveskip=0pt, belowskip=0pt]
s1:  RESUME
s2:  LOAD_CONST 1
     STORE_FAST avar
\end{lstlisting}%
\end{minipage}
\end{center}

The declaration {\tt global avar}
has no runtime representation,
so it contributes no step to the DDG.
A trimmer relying solely on
Definition~\ref{def:debloating} would delete it.
Such a deletion silently changes the semantics,
as the assignment that follows is compiled
to {\tt STORE\_FAST} rather than
{\tt STORE\_GLOBAL}.
This leads to a write to a local slot in {\tt foo}'s frame
instead of the module namespace.
To tackle these cases,
\xtrim\ treats statements involving constructs
that influence compilation
(e.g., {\tt global},
{\tt nonlocal},
{\tt from \_\_future\_\_ import},
{\tt yield},
or {\tt await})
as needed,
regardless of Definition~\ref{def:debloating}.

\point{Value-dependent constructs}
A star import
({\tt from mod import *}) is
unusual in that its semantics depend on a runtime value:
the interpreter reads {\tt mod.\_\_all\_\_}
and looks up every name it lists,
binding each in the importing module.
If the trimmer removes a definition from {\tt mod}
whose name is still included in
{\tt mod.\_\_all\_\_},
the lookup fails
and the star import raises {\tt AttributeError}.
Instead of reasoning about
the contents of {\_\_all\_\_},
our trimmer rewrites a star import
into a regular import as follows.
\begin{lstlisting}[language=Python, basicstyle=\footnotesize\ttfamily]
from .umath import *  # 97 new bindings
from .umath import NAN, PINF, sin  # 3 new bindings 
\end{lstlisting}
It lists exactly the names
on which a surviving step
of the importing module has a data dependence.
With no star import left to consult them,
each {\tt \_\_all\_\_} variable
becomes an ordinary list,
kept or deleted like any other value.

\point{Fallback mechanism}
Our debloating process relies on
observations from executing
the serverless application on the given workloads $W$.
However,
once deployed to the cloud,
the debloated application may receive inputs
other than those in $W$,
and reach code that \xtrim\ removed.
To prevent such failures,
\xtrim\ adopts a fallback mechanism
similar to that of~\ltrim~\cite{l-trim}.
It wraps the serverless function
in an exception handler and,
on failures attributed to the removed code,
falls back to the original application.

\subsection{Implementation Details and Discussion}
\label{sec:implementation}

\xtrim\ is implemented as a command-line tool
using $15$k lines of Python code
and $2$k lines of C code.
The shadow interpreter is pure Python,
built on the {\tt sys.monitoring}~\cite{sys-monitoring} framework,
which hooks every executed bytecode instruction.
The C-API interceptor
(Section~\ref{sec:c-api-interceptor})
is a native extension in C,
and the trimmer uses Python's {\tt ast} module.
We defer the reader to Appendix~\ref{appendix:implementation}
for further details on~\xtrim,
as well as additional optimizations.

\point{Soundness of debloating}
For every workload in $W$
on which the application is deterministic,
$P'$ returns the same value
and performs the same external effects as $P$.
The guarantee relies on Definition~\ref{def:debloating},
which ensures that no step of a surviving statement
depends on a removed one~\cite{korel88dynamic}.

The invariant holds as long as the recorded DDG
over-approximates the true dependences.
The shadow interpreter
(Section~\ref{sec:shadow-interpreter})
sees every executed instruction
and CPython's complete state,
so it captures every dependence carried by bytecode.
The C-API interceptor
(Section~\ref{sec:c-api-interceptor}) records
every access a native extension makes to a Python object,
which the shadow interpreter turns into a dependence.
CPython also exposes macros,
such as {\tt PyList\_GET\_ITEM},
that read an object directly
and leave no call to intercept.
These accesses are covered too,
since a macro operates on a pointer
the extension already holds,
and a pointer can be acquired only as an argument
or through an intercepted function
(e.g., {\tt PyImport\_Import},
{\tt PyObject\_GetAttr}).
We record the dependence
when the pointer is acquired,
on the whole object rather than the element read.
Furthermore,
every external effect
leaves the Python world
through an intercepted {\tt libc} call
(Section~\ref{sec:sinks})
and becomes a sink.
Where the analysis is imprecise,
it retains code rather than removing it.

\point{Limitations}
As a dynamic tool,
\xtrim\ reasons only about the executions it observed.
Inputs beyond $W$,
or non-deterministic behavior,
can reach code that was removed.
Our fallback
(Section~\ref{sec:trimmer})
converts such cases into a re-deployment
of the original application.
This protects correctness rather than latency,
since \ltrim\ measures such a fallback
at 50\,ms of setup plus a second cold start~\cite{l-trim}.
The inputs that trigger it in production
can be added to $W$ for a re-debloating,
and users can also pair~\xtrim{}
with deployment strategies such as canary releases~\cite{canaries}.

\xtrim\ debloats only the Python code of a handler
and its dependencies.
Debloating native extensions would require
combining it with existing binary debloating tools~\cite{razor,binrec}.


\point{Generalizability and extensibility}
\xtrim\ currently supports Python 3.12+.
Because our rules dispatch on bytecode instructions,
porting to a new version requires updating them,
not the design itself.
The underlying concepts
(Section~\ref{sec:approach})
are not specific to Python
and could be applied to languages such as JavaScript,
although the engine would have to be re-implemented
for each execution environment.
We target Python because it is
the dominant serverless runtime
and the source of most production
cold starts~\cite{joosen2025eurosys,datadog-2023}.

\section{Evaluation}
\label{sec:evaluation}

We aim to answer the following questions.

\begin{enumerate}[label={\bf RQ\arabic*}, leftmargin=2.3\parindent]
\item Does~\xtrim{}
preserve the handler's behavior,
and how long does debloating take?
(Section~\ref{sec:eval-time})
\item How effective is
\xtrim{} at reducing cold-start latency
in real-world handlers?
(Section~\ref{sec:eval-effectiveness})
\item How does \xtrim\ affect warm invocations?
(Section~\ref{sec:eval-warm-time})
\item How much do \xtrim's \textit{sink detector}
and \textit{C-API interceptor} contribute to its results?
(Section~\ref{sec:eval-ablation})
\end{enumerate}

\begin{table}[t]
 \footnotesize
  \caption{Success rate and debloating time.
  A tool succeeds on an application when it produces
  a program that returns the answer of the unmodified one.
  The last column is \xtrim's median time
  on the applications the other tool handles.}
  \label{tab:rq1}
  \centering
  \setlength{\tabcolsep}{4pt}
  \begin{tabular}{@{}l r r r r r@{}}
    \toprule
    \textbf{Tool} & \textbf{Successful}
      & \multicolumn{3}{c}{\textbf{Debloating time}}
      & \textbf{\xtrim} \\
    \cmidrule(lr){3-5} \cmidrule(l){6-6}
      & & \textbf{median} & \textbf{min} & \textbf{max}
      & \textbf{median} \\
    \midrule
    \xtrim         & 31/31 & 7\,min & 13\,s & 2.2\,h &  \\
    \ltrim         & 28/31 & 62\,min & 1\,s & 13.8\,h & 5\,min \\
    FaaSLight      & 10/31 & 78\,s & 47\,s & 13\,min & 28\,s \\
    \slimstart     &  7/31 & 35\,s & 10\,s & 57\,s & 27\,s \\
    \bottomrule
  \end{tabular}
\end{table}

\begin{table*}[t]
  \caption{Cold start, initialization and peak memory on AWS Lambda.
  Each tool column gives the median (500 runs) and its change
  against the unmodified application.
  \emph{Share} is the part of the unmodified cold start
  that initialization accounts for.
  Green cells indicate cases where a tool is better than both
  the unmodified application and the other tool
  with statistical significance.
  A $\dagger$ marks a change that is not statistically significant.
  Peak memory counts only above 2\,MB.
  Rows are ordered by the unmodified cold start.}
  \label{tab:rq2}
  \centering
  \footnotesize
  \setlength{\tabcolsep}{2.5pt}
  \begin{tabular}{@{}l rrr rrrr rrr@{}}
    \toprule
    & \multicolumn{3}{c}{\textbf{Cold start (ms)}}
      & \multicolumn{4}{c}{\textbf{Initialization (ms)}}
      & \multicolumn{3}{c}{\textbf{Peak memory (MB)}} \\
    \cmidrule(lr){2-4} \cmidrule(lr){5-8} \cmidrule(l){9-11}
    \textbf{Application}
      & original & \xtrim & \ltrim
      & original & \textit{share} & \xtrim & \ltrim & original & \xtrim & \ltrim \\
    \midrule
    \textsc{ocrmypdf} & 6{,}332 & 6{,}307\,($-0.4$\%)$^{\dagger}$ & 6{,}310\,($-0.3$\%)$^{\dagger}$ & 750 & \textit{12\%} & 744\,($-0.8$\%)$^{\dagger}$ & 745\,($-0.7$\%)$^{\dagger}$ & 190 & \cellcolor{good}{\boldmath\textbf{169\,($-11.1$\%)}} & 185\,($-2.6$\%) \\
    \textsc{huggingface} & 5{,}363 & \cellcolor{good}{\boldmath\textbf{4{,}339\,($-19.1$\%)}} & 4{,}987\,($-7.0$\%) & 4{,}902 & \textit{91\%} & \cellcolor{good}{\boldmath\textbf{3{,}862\,($-21.2$\%)}} & 4{,}489\,($-8.4$\%) & 869 & \cellcolor{good}{\boldmath\textbf{780\,($-10.2$\%)}} & 851\,($-2.1$\%) \\
    \textsc{resnet} & 5{,}149 & \cellcolor{good}{\boldmath\textbf{4{,}249\,($-17.5$\%)}} & \textit{n/a} & 3{,}972 & \textit{77\%} & \cellcolor{good}{\boldmath\textbf{3{,}048\,($-23.3$\%)}} & \textit{n/a} & 955 & \cellcolor{good}{\boldmath\textbf{859\,($-10.1$\%)}} & \textit{n/a} \\
    \textsc{tensorflow} & 4{,}072 & \cellcolor{good}{\boldmath\textbf{3{,}207\,($-21.2$\%)}} & 3{,}519\,($-13.6$\%) & 4{,}066 & \textit{100\%} & \cellcolor{good}{\boldmath\textbf{3{,}200\,($-21.3$\%)}} & 3{,}512\,($-13.6$\%) & 606 & \cellcolor{good}{\boldmath\textbf{501\,($-17.3$\%)}} & 559\,($-7.8$\%) \\
    \textsc{heart-failure} & 3{,}970 & \cellcolor{good}{\boldmath\textbf{2{,}787\,($-29.8$\%)}} & \textit{n/a} & 3{,}060 & \textit{77\%} & \cellcolor{good}{\boldmath\textbf{1{,}878\,($-38.6$\%)}} & \textit{n/a} & 398 & \cellcolor{good}{\boldmath\textbf{265\,($-33.4$\%)}} & \textit{n/a} \\
    \textsc{rnn-generate} & 3{,}065 & 2{,}555\,($-16.7$\%) & \cellcolor{good}{\boldmath\textbf{2{,}399\,($-21.7$\%)}} & 3{,}034 & \textit{99\%} & 2{,}520\,($-17.0$\%) & \cellcolor{good}{\boldmath\textbf{2{,}366\,($-22.0$\%)}} & 621 & \cellcolor{good}{\boldmath\textbf{572\,($-7.9$\%)}} & 576\,($-7.2$\%) \\
    \textsc{ffmpeg} & 2{,}546 & 2{,}521\,($-1.0$\%) & 2{,}515\,($-1.2$\%) & 205 & \textit{8\%} & 176\,($-14.1$\%) & 173\,($-15.8$\%) & 291 & 288\,($-1.0$\%) & 288\,($-1.0$\%) \\
    \textsc{wine} & 2{,}298 & \cellcolor{good}{\boldmath\textbf{1{,}519\,($-33.9$\%)}} & 2{,}049\,($-10.8$\%) & 2{,}271 & \textit{99\%} & \cellcolor{good}{\boldmath\textbf{1{,}497\,($-34.1$\%)}} & 2{,}015\,($-11.3$\%) & 258 & \cellcolor{good}{\boldmath\textbf{164\,($-36.4$\%)}} & 245\,($-5.0$\%) \\
    \textsc{qiskit-nature} & 2{,}257 & \cellcolor{good}{\boldmath\textbf{1{,}529\,($-32.3$\%)}} & 1{,}890\,($-16.3$\%) & 1{,}826 & \textit{81\%} & \cellcolor{good}{\boldmath\textbf{1{,}280\,($-29.9$\%)}} & 1{,}391\,($-23.8$\%) & 288 & \cellcolor{good}{\boldmath\textbf{195\,($-32.3$\%)}} & 258\,($-10.4$\%) \\
    \textsc{sensor-telemetry} & 1{,}992 & \cellcolor{good}{\boldmath\textbf{1{,}258\,($-36.8$\%)}} & \textit{n/a} & 1{,}800 & \textit{90\%} & \cellcolor{good}{\boldmath\textbf{1{,}086\,($-39.7$\%)}} & \textit{n/a} & 214 & \cellcolor{good}{\boldmath\textbf{142\,($-33.6$\%)}} & \textit{n/a} \\
    \textsc{spacy} & 1{,}874 & \cellcolor{good}{\boldmath\textbf{1{,}461\,($-22.0$\%)}} & 1{,}574\,($-16.0$\%) & 1{,}862 & \textit{99\%} & \cellcolor{good}{\boldmath\textbf{1{,}449\,($-22.2$\%)}} & 1{,}562\,($-16.1$\%) & 209 & \cellcolor{good}{\boldmath\textbf{181\,($-13.4$\%)}} & 186\,($-11.0$\%) \\
    \textsc{scikit} & 1{,}774 & \cellcolor{good}{\boldmath\textbf{1{,}327\,($-25.2$\%)}} & 1{,}731\,($-2.4$\%) & 1{,}772 & \textit{100\%} & \cellcolor{good}{\boldmath\textbf{1{,}324\,($-25.2$\%)}} & 1{,}729\,($-2.4$\%) & 187 & \cellcolor{good}{\boldmath\textbf{128\,($-31.6$\%)}} & 182\,($-2.7$\%) \\
    \textsc{sentiment-gzip} & 1{,}656 & \cellcolor{good}{\boldmath\textbf{1{,}144\,($-30.9$\%)}} & 1{,}663\,($+0.4$\%)$^{\dagger}$ & 1{,}651 & \textit{100\%} & \cellcolor{good}{\boldmath\textbf{1{,}139\,($-31.0$\%)}} & 1{,}657\,($+0.4$\%)$^{\dagger}$ & 176 & \cellcolor{good}{\boldmath\textbf{124\,($-29.5$\%)}} & 176\,($+0.0$\%) \\
    \textsc{skimage} & 1{,}546 & \cellcolor{good}{\boldmath\textbf{1{,}044\,($-32.5$\%)}} & 1{,}069\,($-30.9$\%) & 1{,}247 & \textit{81\%} & 793\,($-36.4$\%) & \cellcolor{good}{\boldmath\textbf{771\,($-38.2$\%)}} & 185 & \cellcolor{good}{\boldmath\textbf{137\,($-25.9$\%)}} & 143\,($-22.7$\%) \\
    \textsc{cve-bin-tool} & 1{,}423 & \cellcolor{good}{\boldmath\textbf{1{,}114\,($-21.7$\%)}} & 1{,}140\,($-19.9$\%) & 1{,}107 & \textit{78\%} & \cellcolor{good}{\boldmath\textbf{770\,($-30.4$\%)}} & 827\,($-25.3$\%) & 133 & \cellcolor{good}{\boldmath\textbf{98\,($-26.3$\%)}} & 110\,($-17.3$\%) \\
    \textsc{chdb-olap} & 1{,}170 & 1{,}135\,($-3.0$\%) & 1{,}128\,($-3.6$\%) & 1{,}141 & \textit{97\%} & 1{,}104\,($-3.2$\%) & 1{,}097\,($-3.8$\%) & 304 & 300\,($-1.3$\%) & 300\,($-1.3$\%) \\
    \textsc{jsym} & 856 & \cellcolor{good}{\boldmath\textbf{598\,($-30.2$\%)}} & 816\,($-4.7$\%) & 606 & \textit{71\%} & \cellcolor{good}{\boldmath\textbf{412\,($-31.9$\%)}} & 547\,($-9.7$\%) & 100 & \cellcolor{good}{\boldmath\textbf{58\,($-42.0$\%)}} & 96\,($-4.0$\%) \\
    \textsc{pandas} & 747 & \cellcolor{good}{\boldmath\textbf{571\,($-23.5$\%)}} & 672\,($-10.0$\%) & 703 & \textit{94\%} & \cellcolor{good}{\boldmath\textbf{531\,($-24.4$\%)}} & 615\,($-12.6$\%) & 121 & \cellcolor{good}{\boldmath\textbf{94\,($-22.3$\%)}} & 114\,($-5.8$\%) \\
    \textsc{epub-pdf} & 742 & \cellcolor{good}{\boldmath\textbf{542\,($-27.0$\%)}} & 699\,($-5.9$\%) & 660 & \textit{89\%} & \cellcolor{good}{\boldmath\textbf{471\,($-28.6$\%)}} & 618\,($-6.4$\%) & 106 & \cellcolor{good}{\boldmath\textbf{84\,($-20.8$\%)}} & 99\,($-6.6$\%) \\
    \textsc{lxml} & 589 & 557\,($-5.4$\%) & 552\,($-6.3$\%) & 356 & \textit{60\%} & 327\,($-8.3$\%) & \cellcolor{good}{\boldmath\textbf{319\,($-10.4$\%)}} & 71 & 68\,($-4.2$\%) & \cellcolor{good}{\boldmath\textbf{66\,($-7.0$\%)}} \\
    \textsc{textblob} & 588 & \cellcolor{good}{\boldmath\textbf{445\,($-24.3$\%)}} & 523\,($-11.1$\%) & 481 & \textit{82\%} & \cellcolor{good}{\boldmath\textbf{355\,($-26.1$\%)}} & 432\,($-10.3$\%) & 88 & \cellcolor{good}{\boldmath\textbf{71\,($-19.3$\%)}} & 82\,($-6.8$\%) \\
    \textsc{image-resize} & 579 & \cellcolor{good}{\boldmath\textbf{469\,($-19.0$\%)}} & 556\,($-3.9$\%) & 502 & \textit{87\%} & \cellcolor{good}{\boldmath\textbf{392\,($-21.9$\%)}} & 481\,($-4.2$\%) & 96 & \cellcolor{good}{\boldmath\textbf{84\,($-12.5$\%)}} & 93\,($-3.1$\%) \\
    \textsc{lightgbm} & 558 & 419\,($-25.0$\%) & \cellcolor{good}{\boldmath\textbf{396\,($-29.0$\%)}} & 526 & \textit{94\%} & 379\,($-27.8$\%) & \cellcolor{good}{\boldmath\textbf{362\,($-31.2$\%)}} & 105 & 87\,($-17.1$\%) & \cellcolor{good}{\boldmath\textbf{85\,($-19.0$\%)}} \\
    \textsc{face-detection} & 536 & \cellcolor{good}{\boldmath\textbf{494\,($-7.8$\%)}} & 514\,($-4.1$\%) & 387 & \textit{72\%} & \cellcolor{good}{\boldmath\textbf{344\,($-11.1$\%)}} & 364\,($-6.0$\%) & 97 & \cellcolor{good}{\boldmath\textbf{91\,($-6.2$\%)}} & 95\,($-2.1$\%) \\
    \textsc{dna-visualization} & 370 & \cellcolor{good}{\boldmath\textbf{261\,($-29.5$\%)}} & 279\,($-24.8$\%) & 352 & \textit{95\%} & \cellcolor{good}{\boldmath\textbf{239\,($-32.1$\%)}} & 260\,($-26.1$\%) & 68 & \cellcolor{good}{\boldmath\textbf{53\,($-22.1$\%)}} & 57\,($-16.2$\%) \\
    \textsc{shapely-numpy} & 305 & \cellcolor{good}{\boldmath\textbf{242\,($-20.6$\%)}} & 271\,($-11.0$\%) & 301 & \textit{99\%} & \cellcolor{good}{\boldmath\textbf{236\,($-21.6$\%)}} & 266\,($-11.4$\%) & 61 & \cellcolor{good}{\boldmath\textbf{53\,($-13.1$\%)}} & 58\,($-4.9$\%) \\
    \textsc{110.dynamic-html} & 221 & \cellcolor{good}{\boldmath\textbf{161\,($-27.3$\%)}} & 168\,($-24.0$\%) & 217 & \textit{98\%} & \cellcolor{good}{\boldmath\textbf{156\,($-27.9$\%)}} & 163\,($-24.7$\%) & 50 & 39\,($-22.0$\%) & 40\,($-20.0$\%) \\
    \textsc{igraph} & 209 & \cellcolor{good}{\boldmath\textbf{167\,($-20.1$\%)}} & 175\,($-16.2$\%) & 205 & \textit{98\%} & \cellcolor{good}{\boldmath\textbf{163\,($-20.5$\%)}} & 172\,($-16.3$\%) & 47 & 42\,($-10.6$\%) & 43\,($-8.5$\%) \\
    \textsc{markdown} & 176 & 175\,($-0.7$\%)$^{\dagger}$ & 176\,($+0.3$\%)$^{\dagger}$ & 155 & \textit{88\%} & 153\,($-0.9$\%)$^{\dagger}$ & 155\,($+0.3$\%)$^{\dagger}$ & 40 & 39\,($-2.5$\%) & 40\,($+0.0$\%) \\
    \textsc{encrypt} & 174 & 172\,($-1.5$\%)$^{\dagger}$ & 177\,($+1.5$\%)$^{\dagger}$ & 160 & \textit{92\%} & 161\,($+0.5$\%)$^{\dagger}$ & 161\,($+0.6$\%)$^{\dagger}$ & 46 & 45\,($-2.2$\%) & 46\,($+0.0$\%) \\
    \textsc{compression} & 164 & 158\,($-3.8$\%) & 160\,($-2.6$\%) & 153 & \textit{93\%} & 148\,($-3.6$\%) & 149\,($-3.0$\%) & 43 & 42\,($-2.3$\%) & 42\,($-2.3$\%) \\
    \midrule
    \textbf{Median} &  & $-21.7$\% & $-8.5$\% &  & \textit{91\%} & $-23.3$\% & $-10.9$\% &  & $-17.1$\% & $-5.4$\% \\
    \textbf{Best} &  & \textbf{22}/31 & 2/28 &  &  & \textbf{21}/31 & 4/28 &  & \textbf{22}/31 & 2/28 \\
    \bottomrule
  \end{tabular}
\end{table*}

\subsection{Experimental Setup}
\label{sec:experimental-setup}

\point{Baselines}
For RQs 1--3,
we compare \xtrim\ against the three state-of-the-art tools
discussed in Section~\ref{sec:background}.
\ltrim~\cite{l-trim} removes top-level statements
by delta debugging~\cite{delta-debugging},
FaaSLight~\cite{faaslight} loads
statically unreachable functions on demand,
and \slimstart~\cite{slimstart} defers the imports
of the libraries that cost the most to initialize.

\point{Dataset}
We evaluate on \numberofapps\ serverless applications,
the union of those used by the three tools above,
plus applications from SeBS~\cite{copik2021sebs}
and FunctionBench~\cite{kim2019functionbench}.
We exclude duplicates,
micro-benchmarks,
and applications
that need external infrastructure (e.g., databases).
The set covers a wide range of library use,
including machine learning inference,
document and image processing.
Unmodified,
their cold start latency
ranges from 164\,ms to 6.3\,s.
The benchmarks also provide handler inputs.
\xtrim, \ltrim\ and \slimstart\ analyze each application
on all of its available inputs,
while FaaSLight is static and needs none for its analysis.
More details about our benchmarks
can be found in Appendix~\ref{appendix:provenance}.

\point{Environment Setup}
We run the debloating tools on a local machine,
a 16-core Intel Xeon E5-2650 at 2.30\,GHz
with 40\,GB of memory running Debian~13,
and measure the time spent in analyzing each benchmark.
We then deploy every debloated
application on AWS Lambda
(RQ2 and RQ3),
as a container image
with 3008\,MB of memory and a timeout of 300\,s
in region \texttt{us-east-1},
the same memory and region as \ltrim's artifact.
Since AWS Lambda scales CPU with memory
and allocates one vCPU at~\nnum{1769}\,MB~\cite{aws-lambda-memory},
every run is configured with the same 1.7 vCPUs.
To show that the results are not tied to one platform,
we repeat these experiments on the local machine,
inside the official AWS Lambda Python~3.12 base image
(Appendix~\ref{appendix:local}).

\point{Metrics}
For every application and tool we report the debloating time,
the cold-start and warm-start latency, and the peak memory.
Latency, peak memory,
and the initialization portion of the cold start
all come from AWS's own report on each invocation.
Every metric is the median over 500 invocations
of an application and tool.
We resample these observations
\nnum{10000} times
to obtain 95\% bootstrap confidence intervals
and consider a change statistically significant
when its interval excludes zero.
AWS reports peak memory in whole megabytes
and it varies little across invocations,
so we count a difference only from 2\,MB.
RQ1's success rate counts a tool run as successful
when the debloated program returns the same output
and produces the same side effects
as the unmodified one on every invocation.

\subsection{RQ1: Success Rate and Debloating Time}
\label{sec:eval-time}

\xtrim\ debloats all 31 applications successfully
(Table~\ref{tab:rq1}).
On the other hand,
\ltrim\ succeeds on 28 of them,
producing no debloated artifact for
\textsc{heart-failure},
\textsc{resnet}
and \textsc{sensor-telemetry}.
FaaSLight and \slimstart\ succeed
only on 10 and on 7 applications,
respectively.
Almost all of these failures stem from
implementation defects.

Debloating is an offline cost that is paid once,
but it is not free.
\xtrim\ is faster than \ltrim\
on 27 of the 28 applications it debloats
(median 5 minutes against 62).
This is because
\ltrim\ treats the application
as a black box:
each candidate removal requires
another application run to check
whether the observed behavior is preserved.
FaaSLight and \slimstart\ report lower medians,
78 and 35 seconds,
but only over the 10 and 7 applications they handle,
which are among the cheapest of the dataset.
\xtrim\ needs 28 and 27 seconds on those same applications.
\takeaway{
\xtrim\ is the only tool that debloats
every application successfully,
and it is faster than every other tool.
}

\subsection{RQ2: Effectiveness of~\xtrim}
\label{sec:eval-effectiveness}

Relative to the original application,
\xtrim\ reduces cold-start latency
on 28 of the~\numberofapps\ applications
with statistical significance
(Table~\ref{tab:rq2}).
For half of the 31 applications,
the reduction is more than 21.7\% (median)
and reaches 36.8\% on \textsc{sensor-telemetry},
while no application becomes slower.
The gain comes from initialization,
which \xtrim\ reduces by 23.3\%
and which accounts for a median of 91\%
of the baseline cold start.
\textsc{ocrmypdf} and \textsc{ffmpeg} are the exception,
spending only 12\% and 8\% of their cold start
on initialization.
Regarding peak memory,
the median reduction is 17.1\%.

\ltrim\ reduces the cold start by 8.5\% at the median,
against \xtrim's 21.7\%,
and peak memory follows the same pattern (5.4\% vs. 17.1\%).
Table~\ref{tab:rq2} marks
the applications
for which a tool performs significantly better
than both the original application
and the other tool.
With respect to cold-start time,
\xtrim\ is the best in 22 of the 31,
three of them uncontested since \ltrim\ produced no artifact,
while \ltrim\ is the best in two.
In those two cases,
\ltrim\ wins because \xtrim\ retains
a small number of imports that its dependence analysis
cannot prove unnecessary,
and those imports carry a large
transitive cost.

Neither FaaSLight nor \slimstart\ improves the cold start
of a single application significantly, and are thus omitted
from the table for readability
and report their results in Appendix~\ref{appendix:faaslight-slimstart-cold}.
Cold start latency grows by 16.5\% at the median under FaaSLight,
while \slimstart\ has a minor effect ($+1.3$\%)
on the seven applications it handles.
\takeaway{
\xtrim\ removes about a fifth of the cold start.
That is more than twice as much as the next best tool.
}

\subsection{RQ3: Warm-Start Performance}
\label{sec:eval-warm-time}

\begin{table}[t]
  \caption{Warm invocations. The first column is the median handler time
  of the unmodified application and the rest the change against it.
  Green marks the fastest option and orange a tool slower than
  the unmodified one, both with statistical significance.
  A $\dagger$ marks a change that is not significant.}
  \label{tab:rq3}
  \centering
  \footnotesize
  \setlength{\tabcolsep}{3pt}
  \begin{tabular}{@{}l r rrrr@{}}
    \toprule
    & \textbf{original} & \multicolumn{4}{c}{\textbf{change (\%)}} \\
    \cmidrule(l){3-6}
    \textbf{Application} & \textbf{(ms)} & \textbf{\xtrim} & \textbf{\ltrim} & \textbf{FaaSL.} & \textbf{SlimS.} \\
    \midrule
    \textsc{ocrmypdf} & 5{,}119 & $+0.5$$^{\dagger}$ & $+0.1$$^{\dagger}$ & \textit{n/a} & \textit{n/a} \\
    \textsc{ffmpeg} & 2{,}190 & $+0.4$$^{\dagger}$ & $+0.4$$^{\dagger}$ & \cellcolor{slow}$+1.5$ & $-0.5$$^{\dagger}$ \\
    \textsc{heart-failure} & 781 & \cellcolor{good}{\boldmath\textbf{$-0.7$}} & \textit{n/a} & \textit{n/a} & \textit{n/a} \\
    \textsc{skimage} & 222 & \cellcolor{good}{\boldmath\textbf{$-1.6$}} & $-0.5$$^{\dagger}$ & \textit{n/a} & \textit{n/a} \\
    \textsc{huggingface} & 216 & \cellcolor{good}{\boldmath\textbf{$-4.1$}} & $+0.5$$^{\dagger}$ & \textit{n/a} & \textit{n/a} \\
    \textsc{resnet} & 144 & \cellcolor{good}{\boldmath\textbf{$-4.3$}} & \textit{n/a} & \textit{n/a} & \textit{n/a} \\
    \textsc{sensor-telemetry} & 125 & \cellcolor{good}{\boldmath\textbf{$-4.4$}} & \textit{n/a} & \textit{n/a} & \textit{n/a} \\
    \textsc{face-detection} & 98.6 & $+0.5$$^{\dagger}$ & \cellcolor{slow}$+2.6$ & \cellcolor{good}{\boldmath\textbf{$-1.7$}} & \textit{n/a} \\
    \textsc{qiskit-nature} & 90.5 & \cellcolor{good}{\boldmath\textbf{$-8.1$}} & $+0.3$$^{\dagger}$ & \textit{n/a} & \textit{n/a} \\
    \textsc{cve-bin-tool} & \cellcolor{good}{\boldmath\textbf{74.5}} & \cellcolor{slow}$+1.4$ & \cellcolor{slow}$+0.8$ & \textit{n/a} & \textit{n/a} \\
    \textsc{epub-pdf} & 58.0 & \cellcolor{good}{\boldmath\textbf{$-0.9$}} & $+0.4$$^{\dagger}$ & \textit{n/a} & \textit{n/a} \\
    \textsc{image-resize} & 43.3 & $+0.2$$^{\dagger}$ & $-0.0$$^{\dagger}$ & \textit{n/a} & \textit{n/a} \\
    \textsc{pandas} & 29.9 & \cellcolor{good}{\boldmath\textbf{$-8.5$}} & $-0.5$$^{\dagger}$ & \textit{n/a} & \textit{n/a} \\
    \textsc{chdb-olap} & 28.6 & \cellcolor{slow}$+1.9$ & \cellcolor{slow}$+2.8$ & \cellcolor{slow}$+1.6$ & $+0.0$$^{\dagger}$ \\
    \textsc{lxml} & 21.8 & $-1.4$$^{\dagger}$ & $-1.0$$^{\dagger}$ & \textit{n/a} & $-0.0$$^{\dagger}$ \\
    \textsc{markdown} & 18.2 & $-0.1$$^{\dagger}$ & $+0.3$$^{\dagger}$ & \cellcolor{slow}$+1295.3$ & \cellcolor{slow}$+1.3$ \\
    \textsc{wine} & 15.1 & \cellcolor{good}{\boldmath\textbf{$-8.4$}} & \cellcolor{slow}$+2.5$ & \textit{n/a} & \textit{n/a} \\
    \textsc{dna-visualization} & 14.0 & \cellcolor{slow}$+7.1$ & $+0.2$$^{\dagger}$ & \cellcolor{good}{\boldmath\textbf{$-1.6$}} & \textit{n/a} \\
    \textsc{lightgbm} & 13.0 & $-0.2$$^{\dagger}$ & $-0.5$$^{\dagger}$ & \textit{n/a} & \textit{n/a} \\
    \textsc{jsym} & 12.4 & \cellcolor{good}{\boldmath\textbf{$-4.0$}} & $+0.0$$^{\dagger}$ & \textit{n/a} & \textit{n/a} \\
    \textsc{rnn-generate} & 11.8 & \cellcolor{good}{\boldmath\textbf{$-2.7$}} & $-0.2$$^{\dagger}$ & \textit{n/a} & \textit{n/a} \\
    \textsc{spacy} & 7.57 & \cellcolor{good}{\boldmath\textbf{$-4.4$}} & $-0.9$$^{\dagger}$ & \textit{n/a} & \textit{n/a} \\
    \textsc{compression} & 6.82 & $-0.3$$^{\dagger}$ & $-0.3$$^{\dagger}$ & \cellcolor{slow}$+39.4$ & $-1.2$$^{\dagger}$ \\
    \textsc{textblob} & 4.04 & \cellcolor{good}{\boldmath\textbf{$-4.0$}} & $-1.5$ & \textit{n/a} & \textit{n/a} \\
    \textsc{110.dynamic-html} & 3.37 & \cellcolor{good}{\boldmath\textbf{$-5.6$}} & $+0.0$$^{\dagger}$ & \cellcolor{slow}$+51.3$ & \textit{n/a} \\
    \textsc{tensorflow} & 2.22 & $-0.5$$^{\dagger}$ & $+0.9$$^{\dagger}$ & \textit{n/a} & \textit{n/a} \\
    \textsc{sentiment-gzip} & 2.09 & $-1.0$$^{\dagger}$ & $+0.5$$^{\dagger}$ & \textit{n/a} & \textit{n/a} \\
    \textsc{scikit} & 1.94 & \cellcolor{good}{\boldmath\textbf{$-2.1$}} & $-0.5$$^{\dagger}$ & \textit{n/a} & \textit{n/a} \\
    \textsc{shapely-numpy} & 1.85 & $-1.9$ & $-2.2$ & \cellcolor{slow}$+107.6$ & \textit{n/a} \\
    \textsc{igraph} & 1.78 & $+0.6$$^{\dagger}$ & $+0.6$$^{\dagger}$ & $-0.6$$^{\dagger}$ & $+0.6$$^{\dagger}$ \\
    \textsc{encrypt} & 1.66 & $-0.3$$^{\dagger}$ & $+0.3$$^{\dagger}$ & \cellcolor{slow}$+28.7$ & \cellcolor{slow}$+1.6$ \\
    \midrule
    \textbf{Median} & & $-1.0$ & $+0.2$ & $+15.2$ & $-0.0$ \\
    \textbf{Best} & 1/31 & \textbf{15}/31 & 0/28 & 2/10 & 0/7 \\
    \bottomrule
  \end{tabular}
\end{table}

A warm start invokes the handler directly
in an already initialized execution environment.
Reducing warm-start time is not a focus of~\xtrim.
However,
Table~\ref{tab:rq3} reports the effect
on warm-start time for all tools.
At the median \xtrim\ slightly improves warm starts,
being 1.0\% faster.
Individual applications move in both directions,
with \xtrim\ being the best
on 15 of 31 applications
and slower than the original application on three.
\ltrim\ is neutral, 0.2\% slower at the median,
and slightly slower than the original on four applications.

Upon further inspection on the cases where \xtrim\ is slower
than the baseline, we made an interesting observation
regarding the root cause.
On \textsc{dna-visualization},
its worst case, where execution time increases by 7.1\%,
the trimmed program executes fewer bytecode instructions,
\nnum{163666} against~\nnum{163688},
and uses less memory, which is normally beneficial.
However,
the smaller heap crosses the 128\,KB threshold
at which glibc releases unused memory back to the kernel.
Those pages have to be brought back,
leading to about 280 additional page faults per call.
This case is not a limit
on \xtrim's ability to remove unnecessary code,
and preventing the page faults is an optimization
that applies separately from debloating.

\takeaway{
\xtrim\ slightly improves warm starts for roughly half the applications,
while \ltrim\, and \slimstart\ leave them practically unchanged.
FaaSLight degrades performance.
}

\subsection{RQ4: Ablation Study}
\label{sec:eval-ablation}

\begin{table}[t]
\footnotesize
\caption{Effect of each dataflow-tracking configuration compared against the original applications.
    \emph{Live mod.} is the number of modules loaded at runtime.
    Reductions are averaged over the 12 applications
    where all configurations succeed,
    using 100 local cold starts.}
  \label{tab:rq4}
  \centering
  \setlength{\tabcolsep}{4pt}
  \begin{tabular}{@{}l r rrr@{}}
    \toprule
    \textbf{Tracking} & \textbf{succeeds}
      & \multicolumn{3}{c}{\textbf{change (\%)}} \\
    \cmidrule(lr){3-5}
      & & \textbf{live mod.} & \textbf{cold} & \textbf{init} \\
    \midrule
    Full (\xtrim)   & 31/31 & $-27.8$ & $-22.3$ & $-28.7$ \\
    LOAD is sink                         & 31/31 & $-10.9$ & $-10.5$ & $-14.2$ \\
    No C-API             & 12/31 & $-27.9$ & $-21.1$ & $-28.4$ \\
    \bottomrule
  \end{tabular}
\end{table}

We disable one part of \xtrim's analysis at a time
(Table~\ref{tab:rq4}).
\emph{LOAD is sink} replaces the criterion
of Section~\ref{sec:sinks} with a proxy,
keeping a statement whenever
any other statement reads its result.
\emph{No C-API} keeps the criterion
but disables the interceptor
of Section~\ref{sec:c-api-interceptor}.

Each configuration fails in its own way.
\emph{LOAD is sink} stays safe but removes far less,
because almost every import is read somewhere.
Its reductions roughly halve,
from 27.8\% to 10.9\% for live modules,
from 22.3\% to 10.5\% for the cold start,
and from 28.7\% to 14.2\% for initialization.
Such a criterion cannot remove a statement
together with its dependents
(\textbf{C1}, Section~\ref{sec:background}),
so it keeps {\tt import legacy}
({\tt m3}, Figure~\ref{fig:run-main})
because {\tt m7} reads it.
\emph{No C-API} has the opposite profile.
Disabling the interceptor can only make the slice smaller,
yet it removes almost no extra code,
producing the same module sets as~\xtrim\
on 10 of the 12 applications where it survives.
What it loses is soundness,
succeeding on 12 of 31 applications instead of all of them.
In {\tt numpy},
for instance,
the \texttt{\_multiarray\_umath} extension reads
\texttt{numpy.exceptions.TooHardError}
through \texttt{PyObject\_GetAttr}.
No bytecode performs that read,
so the analysis deletes the definition
at \texttt{numpy/exceptions.py:97},
and every application importing {\tt numpy}
then fails with an \texttt{AttributeError}.

\takeaway{
The \textit{sink detector} is what makes~\xtrim\ effective,
and the \textit{C-API interceptor} is what keeps it safe.
}
\subsection{Discussion}
\label{sec:eval-discussion}

\point{Lazy imports}
Deferring an import is the alternative to removing it,
and PEP~810~\cite{pep8102025} will bring it to Python~3.15
as the \texttt{lazy} keyword on an individual import,
a release scheduled for October 2026~\cite{pep7902025}.
Deferral cannot reduce what the platform downloads,
and it postpones the cost of an import rather than removing it,
so the cost returns on the first invocation that needs the library.
Applying it everywhere is also unsafe.
PEP~690~\cite{pep6902022}, its rejected predecessor,
calls lazy imports a potentially breaking semantic change,
since the side effects of a module are deferred with it,
and warns that libraries break in unexpected ways.
We tested this on Python~3.15.0rc1
by making every eligible import lazy
in each application's dependency tree.
Fifteen of the 29 applications
we could build then failed
before returning a result.
All failures occurred
because deferring an import changed
when a module was initialized,
breaking programs that depended on its initialization side effects
or on a particular module initialization order.
The two techniques are nonetheless complementary,
since \xtrim\ could identify the imports
that are safe to defer.

\point{Threats to validity}
%
%
\xtrim\ removes what the inputs it is given never need,
so a wider set may exercise more code
and leave it less to remove.
We also measure a single cloud configuration,
one region, one memory size and one runtime.
Less memory would mean a smaller CPU share,
which would lengthen initialization and change the gains.
The AWS platform is also noisy.
The reductions we claim are an order of magnitude larger,
and they reproduce on other hardware and off the AWS platform
(Appendix~\ref{appendix:local}).

\section{Related Work}
\label{sec:related-work}

\point{Application-level techniques}
The closest line of work
modifies the application itself,
as an effort to reduce cold-start latency.
This line is represented by
\ltrim, FaaSLight, and \slimstart,
which are described in detail in
Section~\ref{sec:background}.
 
\point{Platform-level cold-start mechanisms}
\label{sec:related:coldstart}
Platform-level work reduces cold-start latency
without touching the application,
by changing how the platform creates and reuses
the execution environments in which a function runs.
Snapshot-and-restore systems skip initialization
by restoring a pre-initialized image,
from gVisor checkpoints~\cite{catalyzer2020}
and VM snapshots~\cite{reap2021,faasnap2022,fireworks2022}
to unikernels~\cite{seuss2020}
and AWS Lambda's SnapStart~\cite{snapstart}.
Provisioning and keep-alive policies instead reduce
how often initialization is paid,
by forking new environments from cached
Zygote containers~\cite{sock2018},
by sharing containers or their layers
across functions~\cite{sand2018,pagurus2022,yu2024rainbowcake},
and by deciding which environments to keep
warm~\cite{faascache2021,icebreaker2022}.
All of these are orthogonal to our work,
which changes the application itself,
and they compose with it,
since a smaller artifact yields a smaller snapshot
and a faster environment load.
\ltrim\ measures an 11\% smaller checkpoint
and up to 42\% off the cost of running with
SnapStart~\cite{l-trim}, and \xtrim\ removes
three times as much peak memory.

\label{subsec:related-debloating}
\point{Software debloating}
Debloating approaches differ in the oracle
that decides what to remove.
Static reachability retains code that may
execute~\cite{jax1999,jshrink2020,stubbifier2022},
and coverage-based approaches retain code
that executes in observed runs~\cite{razor,jdbl2023}.
Test-oracle approaches retain whatever passes a
test~\cite{chisel2018,domgad2020,l-trim}.
\xtrim\ uses a dependence-based oracle,
removing a statement only when no DDG path
connects its steps to a sink.
PyTrim~\cite{pytrim2025} works at the coarser
granularity of declared dependencies.


\label{subsec:related-dift}
\point{Information-flow tracking}
\xtrim's shadow interpreter
falls within the broader class of
dynamic information-flow tracking
systems~\cite{schwartz-dta2010,libdft2012,jalangi2013,ichnaea2020,augur2022},
with labels that are DDG steps
rather than taint tags and with control dependences recorded.
For Python,
DynaPyt~\cite{dynapyt2022}
rewrites source code to insert instrumentation hooks,
while Resin~\cite{resin2009} modifies
the interpreter to check policies at
a boundary that covers every I/O channel.
Unlike this prior work,
\xtrim\ observes bytecode through
\texttt{sys.monitoring}~\cite{pep6692022}
with the application and its
dependencies unmodified.

Tracking dependencies across the Python--C boundary poses
an additional challenge.
PolyCruise~\cite{polycruise2022} requires
native extensions to be rebuilt through
LLVM and tracks only explicit flows.
TruffleTaint~\cite{kreindl2020} requires every language,
including C,
to execute under interpretation on GraalVM~\cite{graalvm2013}.
\xtrim's C-API interceptor instead recovers
native dependences from prebuilt wheels.

\section{Conclusion}
\label{sec:conclusion}
Cold-start latency is a major concern in serverless computing,
yet much of what a handler runs
before serving a request never affects its behavior.
\xtrim\ removes that code,
slicing across data and control dependences
in Python and native code alike.
On~\numberofapps\ applications it cuts cold-start latency
by~\empirical{21.7\%} and peak memory by~\empirical{17.1\%}
at the median,
more than twice the best prior tool,
showing that dynamic slicing is an effective basis
for optimizing serverless programs.
The dependence analysis behind~\xtrim\ is not specific
to debloating, and could serve other purposes,
such as source-sink vulnerability detection.

\bibliographystyle{ACM-Reference-Format}
\bibliography{main}

\appendix

\begin{table*}[!t]
  \caption{Effect of representative instructions.
    The rules assume execution
    in the top-level code of a module $\shd{mod}$,
    which affects only {\tt STORE\_NAME}:
    inside a function frame it writes to the shadow store alone
    and not to the heap,
    since the binding does not escape.
    $\ell_{v}$ is the label of value $v$,
    while bare $v$ is the real object,
    read from CPython's operand stack.
    The helper $\shd{id}(\cdot)$ gives an object's address $\alpha$,
    $\shd{store}[n]$ looks up the name $n$ in the shadow store, 
    and $\shd{heap}[\alpha]$ looks up the address $\alpha$ in the shadow heap.}
  \label{tab:shadow-rules}
  \centering
  \resizebox{\linewidth}{!}{
  \begin{tabular}{@{}l >{\ttfamily}l l l l l@{}}
    \toprule
    \textbf{Category}
      & \multicolumn{1}{l}{\textbf{Instruction}}
      & \textbf{Shadow frame} & \textbf{Shadow heap}
      & \textbf{Control context}
      & \textbf{Dependences recorded} \\
    \midrule
    \multirow{2}{*}{Source}
      & IMPORT\_NAME $n$ & push $s$ & -- & push $s$ & -- \\
      & MAKE\_FUNCTION & push $s$ & -- & -- & -- \\
    \midrule
    \multirow{2}{*}{Locals}
      & LOAD\_NAME $n$ & push $\shd{store}[n]$ & -- & --
        & $s \dedge \shd{store}[n]$ \\
      & STORE\_NAME $n$
        & pop $\ell_{\mathit{val}}$;\enspace
          $\shd{store}[n] \leftarrow \ell_{\mathit{val}}$
        & $\shd{heap}[\shd{id}(\shd{mod}.n)] \leftarrow
          \ell_{\mathit{val}}$
        & --
        & $s \dedge \ell_{\mathit{val}}$;\enspace
          $\mathit{chain}(s)$ \\
    \midrule
    \multirow{2}{*}{Heap}
      & LOAD\_ATTR $n$
        & pop $\ell_{\mathit{obj}}$;\enspace
          push $\shd{heap}[\shd{id}(\mathit{obj}.n)]$
        & -- & --
        & $s \dedge \ell_{\mathit{obj}},\
          \shd{heap}[\shd{id}(\mathit{obj}.n)]$ \\
      & STORE\_ATTR $n$ & pop $\ell_{\mathit{val}}, \ell_{\mathit{obj}}$
        & $\shd{heap}[\shd{id}(\mathit{obj}.n)] \leftarrow s$
        & --
        & $s \dedge \ell_{\mathit{val}}, \ell_{\mathit{obj}}$;\enspace
          $\mathit{chain}(s)$ \\
    \midrule
    Compute
      & BINARY\_OP & pop $\ell_a, \ell_b$;\enspace push $s$ & -- & --
        & $s \dedge \ell_a, \ell_b$ \\
    \midrule
    Jump
      & POP\_JUMP\_IF\_* & pop $\ell_{\mathit{pred}}$ & --
        & push $\ell_{\mathit{pred}}$
        & $s \dedge \ell_{\mathit{pred}}$ \\
    \midrule
    \multirow{2}{*}{Function}
      & CALL & pop $\ell_{\mathit{fn}}, \ell_{\mathit{args}}$;\enspace
        new frame with $\shd{store} \leftarrow \ell_{\mathit{args}}$
        & -- & push $s$
        & $s \dedge \ell_{\mathit{fn}}, \ell_{\mathit{args}}$ \\
      & RETURN\_VALUE & pop $\ell_{\mathit{ret}}$;\enspace drop frame;\enspace
        push $\ell_{\mathit{ret}}$
        & -- & pop
        & $s \dedge \ell_{\mathit{ret}}$ \\
    \bottomrule
    \end{tabular}}
\end{table*}

\section{Shadow Interpreter Rules}
\label{appendix:rules}

Table~\ref{tab:shadow-rules} gives the rules
the shadow interpreter applies
for representative instructions
using the notation introduced in Section~\ref{sec:shadow-interpreter}.
{\tt IMPORT\_NAME} pushes the step
that enters a module's top-level code,
{\tt CALL} the step that enters a function frame,
and the conditional jumps
the label of the predicate under test,
which stays on the context
while the branch is in effect.
{\tt RETURN\_VALUE} is the one rule shown that pops.
The ten instructions in the table stand for
the 122 opcodes for which
the engine registers a rule.
The remaining ones neither move labeled values
nor affect control,
and a default rule keeps the shadow stack
aligned with the real one
using the instruction's declared stack effect.

\begin{table*}[!t]
  \caption{Representative transfer rules for CPython built-in calls.
    Notation follows Table~\ref{tab:shadow-rules}:
    $\ell_v$ is the label set of value $v$,
    $\shd{heap}[\shd{id}(o.n)]$ the shadow-heap cell for field $n$ of
    object $o$, and $s$ the executing shadow step.
    We write $\shd{w}(o) = \shd{heap}[\shd{id}(o.{*})]$
    for the whole-object cell of $o$.
    A dash means the rule declares nothing for that column,
    so the default applies;
    $\emptyset$ means an explicitly empty set.}
  \label{tab:native-summaries}
  \centering
  \small
  \setlength{\tabcolsep}{4pt}
  \begin{tabularx}{\textwidth}{@{}>{\ttfamily}l Y Y Y Y@{}}
    \toprule
    \multicolumn{1}{@{}l}{\textbf{Callee}}
      & \textbf{Read scope}
      & \textbf{Shadow heap}
      & \textbf{Result}
      & \textbf{Dependences recorded} \\
    \midrule

    setattr($o,n,v$)
      & --
      & $\shd{heap}[\shd{id}(o.n)] \leftarrow \ell_v$
      & $\emptyset$
      & $\mathit{chain}(s)$ \\

    getattr($o,n,d$)
      & $\shd{heap}[\shd{id}(o.n)]$;\enspace $\shd{w}(o)$
      & --
      & $\ell_o \cup \shd{heap}[\shd{id}(o.n)]
         \cup \shd{w}(o) \cup \ell_d$
      & $s \dedge \shd{heap}[\shd{id}(o.n)],\ \shd{w}(o)$ \\

    dict.get($d,k,x$)
      & $\shd{heap}[\shd{id}(d.k)]$;\enspace $\shd{w}(d)$
      & --
      & $\ell_d \cup \shd{heap}[\shd{id}(d.k)]
         \cup \shd{w}(d) \cup \ell_x$
      & $s \dedge \shd{heap}[\shd{id}(d.k)],\ \shd{w}(d)$ \\

    dict.update($d,o$)
      & $\shd{heap}[\shd{id}(o.k)]$ for $k \in o$;\enspace $\shd{w}(o)$
      & $\shd{heap}[\shd{id}(d.k)] \leftarrow
         \shd{heap}[\shd{id}(o.k)] \cup \ell_o$
      & $\emptyset$
      & $\mathit{chain}(s)$ on written cells \\

    list.append($l,v$)
      & --
      & $\shd{w}(l) \leftarrow \ell_v$
      & $\emptyset$
      & $s \dedge \shd{w}(l)$ \\

    set.add($t,v$)
      & --
      & $\shd{w}(t) \leftarrow \ell_v$
      & $\emptyset$
      & $s \dedge \shd{w}(t)$ \\

    len($x$)
      & $\shd{w}(x)$
      & --
      & $\ell_x$
      & $s \dedge \shd{w}(x)$ \\

    \bottomrule
  \end{tabularx}
\end{table*}

\section{Transfer Rules for CPython Built-in Callees}
\label{appendix:transfer-rules}

CPython's built-in functions and methods of built-in types execute inside the
interpreter rather than through the C API interception layer.  The shadow
interpreter therefore models their effects using transfer rules derived from
the Python 3.12 documentation and, where necessary, the CPython interpreter
source.  The rules specify which shadow-heap cells are read or written and
how the labels of returned values are derived.  Table~\ref{tab:native-summaries}
lists representative rules. The complete table is generated from the
registered summaries.

The rules are applied at the call site.
Each rule declares its read scope
per operand;
in the absence of a narrower declaration,
an operand is treated
as a whole-object read.
For example, \texttt{len(x)} does not read
the interior of $x$,
while \texttt{dict.get} reads
the requested mapping cell
and the mapping's whole-object cell.
Writes similarly identify the
whole-object or field cell that is actually mutated.
When a call can match multiple rules,
their outcomes are conservatively unioned.

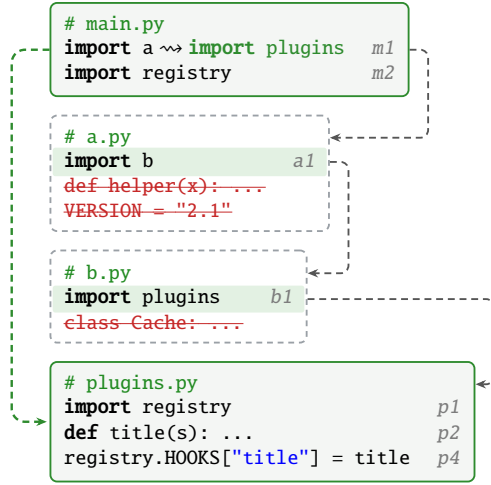
\begin{figure}[t]
  \centering
\usetikzlibrary{arrows.meta,calc,matrix,fit,backgrounds}%
\definecolor{keepline}{HTML}{2E8B2E}%
\definecolor{keepfill}{HTML}{F4F6F4}%
\definecolor{passline}{HTML}{9AA0A6}%
\definecolor{steplab}{HTML}{7C7C7C}%
\definecolor{edgecol}{HTML}{555555}%
\definecolor{delline}{HTML}{C62828}%
\definecolor{needfill}{HTML}{E3F1E3}%
\definecolor{lstcmt}{rgb}{0.133,0.545,0.133}%
\definecolor{lststr}{rgb}{0,0,1}%
\providecommand{\icfn}[1]{\textcolor{lstcmt}{\# #1}}%
\providecommand{\icst}[1]{\textcolor{steplab}{\textit{#1}}}%
\providecommand{\icrm}[1]{\textcolor{delline}{\st{#1}}}%
\providecommand{\ickw}[1]{\textbf{#1}}%
\providecommand{\icstr}[1]{\textcolor{lststr}{#1}}%
\providecommand{\icband}[2]{\fill[needfill, rounded corners=1pt]
  ($(#1.west|-#2.north)+(1.2pt,1.8pt)$) rectangle ($(#1.east|-#2.south)-(1.2pt,1.8pt)$);}%
\begin{tikzpicture}[
  mm/.style={matrix of nodes, ampersand replacement=\&, inner sep=0pt,
             nodes={anchor=base west, inner sep=0pt, font=\footnotesize\ttfamily},
             row sep=2.6pt, column sep=10pt, anchor=north west},
  mod/.style={rectangle, rounded corners=2.5pt, draw=keepline, fill=keepfill,
              line width=0.7pt, inner sep=5pt},
  pass/.style={mod, draw=passline, dashed, dash pattern=on 2.2pt off 1.8pt, fill=white},
  imp/.style={-{Stealth[length=4pt,width=3.2pt]}, line width=0.7pt, draw=edgecol,
        dashed, dash pattern=on 2.6pt off 1.8pt},
  new/.style={imp, draw=keepline, line width=0.9pt},
  note/.style={font=\scriptsize\itshape, text=passline, inner sep=1.5pt},
  elab/.style={font=\scriptsize\itshape, text=steplab, inner sep=2pt},
]
\def\icvs{0.42}

\matrix (MM) [mm] at (0,0) {
  \icfn{main.py} \& \\
  \ickw{import} a\;$\rightsquigarrow$\;\textcolor{keepline}{\ickw{import} plugins} \& \icst{m1} \\
  \ickw{import} registry \& \icst{m2} \\ };
\begin{scope}[on background layer]\node[mod, fit=(MM)] (m) {};\end{scope}

\matrix (AA) [mm] at ($(m.south -| MM.west)+(0,-\icvs)$) {
  \icfn{a.py} \& \\
  \ickw{import} b \& \icst{a1} \\
  \icrm{def helper(x): ...} \& \\
  \icrm{VERSION = "2.1"} \& \\ };
\begin{scope}[on background layer]\node[pass, fit=(AA)] (a) {};\end{scope}

\matrix (BB) [mm] at ($(a.south -| MM.west)+(0,-\icvs)$) {
  \icfn{b.py} \& \\
  \ickw{import} plugins \& \icst{b1} \\
  \icrm{class Cache: ...} \& \\ };
\begin{scope}[on background layer]\node[pass, fit=(BB)] (b) {};\end{scope}

\matrix (PP) [mm] at ($(b.south -| MM.west)+(0,-\icvs)$) {
  \icfn{plugins.py} \& \\
  \ickw{import} registry \& \icst{p1} \\
  \ickw{def} title(s): ... \& \icst{p2} \\
  registry.HOOKS[\icstr{"title"}] = title \& \icst{p4} \\ };
\begin{scope}[on background layer]\node[mod, fit=(PP)] (p) {};\end{scope}

\begin{scope}[on background layer]
  \icband{a}{AA-2-1}
  \icband{b}{BB-2-1}
\end{scope}

\newcommand{\icedge}[4]{
  \draw[imp, rounded corners=3pt]
    let \p1=(#1.east|-#2), \p2=(#3.east|-#4), \n1={max(\x1,\x2)+7pt} in
    (\p1) -- (\n1,\y1) -- (\n1,\y2) -- (\p2);}
\icedge{m}{MM-2-1}{a}{AA-1-1}
\icedge{a}{AA-2-1}{b}{BB-1-1}
\icedge{b}{BB-2-1}{p}{PP-1-1}

\coordinate (iclane) at ($(m.west)+(-0.5,0)$);
\draw[new, rounded corners=5pt]
  (m.west |- MM-2-1) -- (iclane |- MM-2-1) -- (iclane |- p.west) -- ([xshift=-1pt]p.west);
\end{tikzpicture}
  \caption{Shortening redundant import chains.
  Arrows denote the import direction,
  i.e., the~\emph{reverse} of the control-dependence edges
  of Figure~\ref{fig:ddg}.}
  \label{fig:collapse}
\end{figure}

\section{Implementation Details}
\label{appendix:implementation}

\point{Accessing CPython's operand stack}
Although {\tt sys.monitoring} exposes
information about the call stack,
each instruction's source location,
and the locals of the executing frame,
it does not expose CPython's operand stack.
\xtrim\ adds a C extension
that peeks the top $n$ values
from the operand stack.
Access to the real operands lets
the engine resolve dynamic accesses.
For example,
a {\tt getattr(obj, name)} call,
whose attribute is computed at runtime,
is handled similarly to a static attribute access
(via {\tt LOAD\_ATTR}),
and the same holds for {\tt eval}
and dynamic imports.

\point{Redundant import chains}
Side-effecting imports
(Section~\ref{sec:background})
force the trimmer to keep import statements
that no data dependence points to.
The module performing
the side effect must be loaded,
but reaching it may require loading
a chain of intermediates that contribute nothing else,
and are paid for on every cold start.

Figure~\ref{fig:collapse} shows such a case,
a variant of our running example
in which {\tt main.py} reaches
the side-effecting module {\tt plugins}
(recall the update at {\tt p4},
Figure~\ref{fig:run-modules})
not directly,
but through {\tt a} and then {\tt b}.
Here,
{\tt main.py} imports {\tt a},
but does not directly access
any of its contents.
Therefore,
the steps {\tt m1},
{\tt a1},
and {\tt b1} survive
only because they lie on the path
that triggers {\tt p4}.
Notably,
nothing else in {\tt a} or {\tt b} is needed
(red lines in Figure~\ref{fig:collapse}),
yet both are still loaded on every cold start.

Our trimmer addresses
this redundant chain of import statements
with a rewrite
we call~\textit{import collapsing},
which is applied while parsing each module.
When the trimmer reaches a surviving import statement,
it first checks the DDG for a data dependence
on the imported module.
A data dependence indicates
that a surviving step in the importing module
reads a name that the import statement binds.
In such a case,
the import statement stays as written.
Otherwise the import is kept
only for the side effect it triggers,
and the trimmer looks for
where that effect actually lives.
Since an import step controls
every top-level step of the module it loaded
(Figure~\ref{fig:ddg}),
its outgoing control edges lead to
that module,
one hop at a time.
The walk stops as soon as it reaches
a module with more than one surviving step.
The trimmer then rewrites
the original import to load
the module where the walk stopped.

In Figure~\ref{fig:collapse},
the statement {\tt import a} at {\tt m1} is kept
but no step in {\tt main.py}
reads the name {\tt a}.
Therefore,
the walk follows the control
dependences from {\tt m1} to {\tt a1},
the only surviving step of {\tt a},
then to {\tt b1},
the only surviving step of {\tt b},
and reaches {\tt plugins},
which has more than one needed step.
So {\tt import a} becomes {\tt import plugins},
and neither {\tt a} nor {\tt b} is ever loaded.

\begin{table*}[!t]
  \centering
  \caption{The \numberofapps\ benchmark applications, their provenance, and their footprint. 
  \emph{Introduced by} is the suite or paper that first published the application; 
  \emph{Used by} lists the prior debloaters/frameworks evaluated on it.
  All \numberofapps\ are evaluated by \xtrim. 
  \emph{Size (MB)} is the total size of the installed dependency tree, 
  \emph{Deps} is the count of package dependencies, 
  \emph{Imports} is the static number of \texttt{import} statements in that tree, and 
  \emph{Live} is the number of modules actually loaded at runtime.
Suites: 
  FaaSLight~\cite{faaslight}, 
  FunctionBench~\cite{kim2019functionbench},
  $\lambda$-trim~\cite{l-trim}, 
  RainbowCake~\cite{yu2024rainbowcake}, 
  SeBS~\cite{copik2021sebs}, 
  SlimStart~\cite{slimstart}.
  }
  \label{tab:app-provenance}
  \small
  \setlength{\tabcolsep}{5pt}
  \begin{tabular}{@{}lll rrrr@{}}
    \toprule
    Application & Introduced by & Used by & Size (MB) & Deps & Imports & Live \\
    \midrule
    110.dynamic-html   & SeBS          & SeBS                                        & 51.19    & 14 & 11,529  & 99   \\
    chdb-olap          & $\lambda$-trim & $\lambda$-trim                              & 632.53   & 15 & 11,252  & 82   \\
    compression        & SeBS          & RainbowCake, $\lambda$-trim                 & 54.73    & 17 & 11,214  & 11   \\
    cve-bin-tool       & SlimStart     & SlimStart                                   & 277.48   & 94 & 50,389  & 1,393 \\
    dna-visualization  & SeBS          & SeBS, RainbowCake, $\lambda$-trim, SlimStart & 284.14   & 51 & 29,243  & 226  \\
    encrypt            & $\lambda$-trim & $\lambda$-trim                              & 63.85    & 19 & 11,972  & 62   \\
    epub-pdf           & $\lambda$-trim & $\lambda$-trim                              & 95.04    & 27 & 16,169  & 645  \\
    face-detection     & FunctionBench & FunctionBench                               & 288.75   & 14 & 15,833  & 200  \\
    ffmpeg             & SeBS          & SeBS, RainbowCake, $\lambda$-trim           & 53.91    & 16 & 12,788  & 85   \\
    heart-failure      & SlimStart     & SlimStart                                   & 1,582.72 & 41 & 81,845  & 1,898 \\
    huggingface        & FaaSLight     & FaaSLight, $\lambda$-trim                   & 5,037.33 & 66 & 115,541 & 2,095 \\
    igraph             & SeBS          & SeBS, RainbowCake, $\lambda$-trim           & 59.53    & 17 & 11,818  & 129  \\
    image-resize       & FaaSLight     & FaaSLight, $\lambda$-trim                   & 49.39    & 15 & 11,287  & 342  \\
    jsym               & $\lambda$-trim & $\lambda$-trim                              & 124.26   & 16 & 45,898  & 602  \\
    lightgbm           & FaaSLight     & FaaSLight, $\lambda$-trim                   & 304.09   & 19 & 27,877  & 380  \\
    lxml               & FaaSLight     & FaaSLight, $\lambda$-trim                   & 64.99    & 23 & 11,842  & 292  \\
    markdown           & RainbowCake   & RainbowCake, $\lambda$-trim                 & 50.84    & 15 & 11,363  & 66   \\
    ocrmypdf           & SlimStart     & SlimStart                                   & 191.65   & 45 & 28,380  & 865  \\
    pandas             & $\lambda$-trim & $\lambda$-trim                              & 191.57   & 20 & 33,721  & 543  \\
    qiskit-nature      & $\lambda$-trim & $\lambda$-trim                              & 698.62   & 33 & 77,797  & 2,244 \\
    resnet             & SeBS          & SeBS, RainbowCake, $\lambda$-trim           & 4,892.75 & 46 & 83,675  & 1,999 \\
    rnn-generate       & FunctionBench & FunctionBench                               & 4,902.96 & 42 & 88,857  & 1,064 \\
    scikit             & FaaSLight     & FaaSLight, $\lambda$-trim                   & 409.65   & 22 & 37,710  & 1,047 \\
    sensor-telemetry   & SlimStart     & SlimStart                                   & 416.33   & 35 & 58,231  & 1,048 \\
    sentiment-gzip     & $\lambda$-trim & $\lambda$-trim                              & 410.29   & 23 & 37,726  & 980  \\
    shapely-numpy      & $\lambda$-trim & $\lambda$-trim                              & 127.08   & 16 & 16,915  & 169  \\
    skimage            & FaaSLight     & FaaSLight, $\lambda$-trim                   & 448.17   & 33 & 34,204  & 950  \\
    spacy              & $\lambda$-trim & $\lambda$-trim                              & 231.56   & 58 & 29,425  & 906  \\
    tensorflow         & FaaSLight     & FaaSLight, $\lambda$-trim                   & 2,051.63 & 65 & 64,142  & 2,906 \\
    textblob           & RainbowCake   & RainbowCake, $\lambda$-trim                 & 74.45    & 28 & 16,434  & 431  \\
    wine               & FaaSLight     & FaaSLight, $\lambda$-trim                   & 422.42   & 65 & 55,520  & 1,598 \\
    \bottomrule
  \end{tabular}
\end{table*}

\section{Benchmark Provenance and Characteristics}
\label{appendix:provenance}

Table~\ref{tab:app-provenance} details the provenance,
dependency counts, and sizes of the \numberofapps\ benchmark applications
introduced in Section~\ref{sec:experimental-setup}.
For each application,
it reports the benchmark suite or paper that introduced it,
the prior debloaters evaluated on it,
the total size of the installed dependency tree,
the number of package dependencies,
the static count of \texttt{import} statements across those dependencies,
and the number of modules loaded at runtime.

\section{Measurements on the Local Machine}
\label{appendix:local}

We repeat the campaign off the platform,
inside the same Lambda base image
on the machine of Section~\ref{sec:experimental-setup},
with two CPUs and 3008\,MB.
Every application and tool runs
50 cold and 300 warm times.
Table~\ref{tab:local-cold} reports the cold start,
while Table~\ref{tab:local-warm} the warm invocations.
\xtrim\ takes 24.7\% off the cold start at the median
and \ltrim\ 8.2\%,
while FaaSLight adds 37.0\% and \slimstart\ 3.0\%.
Warm invocations move as little as they do on AWS,
by $-1.2$\% for \xtrim\ and $+0.6$\% for \ltrim.

\clearpage
\begin{table}[t]
  \caption{Cold start on the local machine, the module-level code plus the handler, over 50 runs of each application and tool. The first column is the median of the unmodified application and the rest the change against it. A $\dagger$ marks a change that is not statistically significant.}
  \label{tab:local-cold}
  \centering
  \footnotesize
  \setlength{\tabcolsep}{3pt}
  \begin{tabular}{@{}l r rrrr@{}}
    \toprule
    & \textbf{original} & \multicolumn{4}{c}{\textbf{change (\%)}} \\
    \cmidrule(l){3-6}
    \textbf{Application} & \textbf{(ms)} & \textbf{\xtrim} & \textbf{\ltrim} & \textbf{FaaSL.} & \textbf{SlimS.} \\
    \midrule
    \textsc{huggingface} & 7{,}752 & $-17.6$ & $-8.2$ & \textit{n/a} & \textit{n/a} \\
    \textsc{ocrmypdf} & 6{,}738 & $-3.7$ & $+1.0$$^{\dagger}$ & \textit{n/a} & \textit{n/a} \\
    \textsc{resnet} & 6{,}014 & $-18.4$ & \textit{n/a} & \textit{n/a} & \textit{n/a} \\
    \textsc{ffmpeg} & 3{,}606 & $-3.9$$^{\dagger}$ & $-0.5$$^{\dagger}$ & $-7.8$ & $+9.3$ \\
    \textsc{heart-failure} & 3{,}356 & $-24.7$ & \textit{n/a} & \textit{n/a} & \textit{n/a} \\
    \textsc{tensorflow} & 3{,}324 & $-25.2$ & $-18.7$ & \textit{n/a} & \textit{n/a} \\
    \textsc{rnn-generate} & 2{,}443 & $-26.4$ & $-36.6$ & \textit{n/a} & \textit{n/a} \\
    \textsc{qiskit-nature} & 2{,}288 & $-26.6$ & $-13.6$ & \textit{n/a} & \textit{n/a} \\
    \textsc{wine} & 2{,}283 & $-32.4$ & $-11.4$ & \textit{n/a} & \textit{n/a} \\
    \textsc{spacy} & 1{,}772 & $-23.5$ & $-14.3$ & \textit{n/a} & \textit{n/a} \\
    \textsc{scikit} & 1{,}690 & $-25.8$ & $-1.2$$^{\dagger}$ & \textit{n/a} & \textit{n/a} \\
    \textsc{sentiment-gzip} & 1{,}652 & $-26.8$ & $-1.0$$^{\dagger}$ & \textit{n/a} & \textit{n/a} \\
    \textsc{sensor-telemetry} & 1{,}645 & $-26.2$ & \textit{n/a} & \textit{n/a} & \textit{n/a} \\
    \textsc{skimage} & 1{,}474 & $-25.1$ & $-18.8$ & \textit{n/a} & \textit{n/a} \\
    \textsc{cve-bin-tool} & 1{,}187 & $-22.9$ & $-19.1$ & \textit{n/a} & \textit{n/a} \\
    \textsc{chdb-olap} & 952 & $-11.5$$^{\dagger}$ & $-7.2$$^{\dagger}$ & $-8.0$$^{\dagger}$ & $-17.4$ \\
    \textsc{jsym} & 714 & $-28.0$ & $-5.2$ & \textit{n/a} & \textit{n/a} \\
    \textsc{face-detection} & 707 & $-2.5$$^{\dagger}$ & $-2.3$$^{\dagger}$ & $+107.4$ & \textit{n/a} \\
    \textsc{pandas} & 645 & $-18.6$ & $-8.3$ & \textit{n/a} & \textit{n/a} \\
    \textsc{epub-pdf} & 559 & $-33.9$ & $-15.3$ & \textit{n/a} & \textit{n/a} \\
    \textsc{lightgbm} & 524 & $-29.0$ & $-30.9$ & \textit{n/a} & \textit{n/a} \\
    \textsc{lxml} & 463 & $-7.1$ & $-6.7$ & \textit{n/a} & $+0.8$$^{\dagger}$ \\
    \textsc{image-resize} & 436 & $-26.9$ & $-7.5$ & \textit{n/a} & \textit{n/a} \\
    \textsc{textblob} & 418 & $-29.3$ & $-13.5$ & \textit{n/a} & \textit{n/a} \\
    \textsc{dna-visualization} & 335 & $-26.1$ & $-25.2$ & $+33.1$ & \textit{n/a} \\
    \textsc{shapely-numpy} & 291 & $-12.4$ & $-6.7$ & $+40.9$ & \textit{n/a} \\
    \textsc{110.dynamic-html} & 89.4 & $-46.1$ & $-41.4$ & $+21.7$ & \textit{n/a} \\
    \textsc{igraph} & 89.0 & $-16.8$ & $-16.0$ & $+57.0$ & $+3.0$$^{\dagger}$ \\
    \textsc{markdown} & 70.2 & $-8.2$$^{\dagger}$ & $-8.0$ & $+463.7$ & $+4.7$$^{\dagger}$ \\
    \textsc{encrypt} & 46.0 & $-8.9$$^{\dagger}$ & $+1.4$$^{\dagger}$ & $+43.3$ & $+5.7$$^{\dagger}$ \\
    \textsc{compression} & 31.9 & $+0.5$$^{\dagger}$ & $+5.6$$^{\dagger}$ & $+27.7$ & $-0.8$$^{\dagger}$ \\
    \midrule
    \textbf{Median} &  & $-24.7$ & $-8.2$ & $+37.0$ & $+3.0$ \\
    \textit{applications} &  & \textit{31} & \textit{28} & \textit{10} & \textit{7} \\
    \bottomrule
  \end{tabular}
\end{table}

\begin{table}[t]
  \caption{Warm invocations on the local machine, over 300 runs of each application and tool. The first column is the median handler time of the unmodified application and the rest the change against it. A $\dagger$ marks a change that is not statistically significant.}
  \label{tab:local-warm}
  \centering
  \footnotesize
  \setlength{\tabcolsep}{3pt}
  \begin{tabular}{@{}l r rrrr@{}}
    \toprule
    & \textbf{original} & \multicolumn{4}{c}{\textbf{change (\%)}} \\
    \cmidrule(l){3-6}
    \textbf{Application} & \textbf{(ms)} & \textbf{\xtrim} & \textbf{\ltrim} & \textbf{FaaSL.} & \textbf{SlimS.} \\
    \midrule
    \textsc{ocrmypdf} & 5{,}940 & $-1.5$ & $-1.4$ & \textit{n/a} & \textit{n/a} \\
    \textsc{ffmpeg} & 3{,}407 & $-1.3$$^{\dagger}$ & $+1.4$$^{\dagger}$ & $-1.8$$^{\dagger}$ & $+8.3$ \\
    \textsc{huggingface} & 2{,}299 & $-4.3$ & $-4.3$ & \textit{n/a} & \textit{n/a} \\
    \textsc{resnet} & 1{,}094 & $-16.9$ & \textit{n/a} & \textit{n/a} & \textit{n/a} \\
    \textsc{heart-failure} & 1{,}087 & $-3.5$ & \textit{n/a} & \textit{n/a} & \textit{n/a} \\
    \textsc{qiskit-nature} & 416 & $-2.3$$^{\dagger}$ & $-1.1$$^{\dagger}$ & \textit{n/a} & \textit{n/a} \\
    \textsc{lxml} & 248 & $-0.2$$^{\dagger}$ & $-0.0$$^{\dagger}$ & \textit{n/a} & $+0.5$$^{\dagger}$ \\
    \textsc{skimage} & 237 & $+0.0$$^{\dagger}$ & $-1.6$ & \textit{n/a} & \textit{n/a} \\
    \textsc{sensor-telemetry} & 170 & $+0.9$$^{\dagger}$ & \textit{n/a} & \textit{n/a} & \textit{n/a} \\
    \textsc{face-detection} & 117 & $-2.8$ & $-0.7$$^{\dagger}$ & $-2.4$ & \textit{n/a} \\
    \textsc{cve-bin-tool} & 112 & $-3.4$ & $+20.3$ & \textit{n/a} & \textit{n/a} \\
    \textsc{rnn-generate} & 101 & $-1.0$ & $+0.1$$^{\dagger}$ & \textit{n/a} & \textit{n/a} \\
    \textsc{image-resize} & 56.8 & $+5.0$ & $+2.4$ & \textit{n/a} & \textit{n/a} \\
    \textsc{epub-pdf} & 55.5 & $+12.1$ & $+4.7$$^{\dagger}$ & \textit{n/a} & \textit{n/a} \\
    \textsc{chdb-olap} & 42.4 & $-0.6$$^{\dagger}$ & $-1.2$$^{\dagger}$ & $-1.3$$^{\dagger}$ & $-7.1$ \\
    \textsc{pandas} & 37.7 & $-8.8$ & $+1.1$ & \textit{n/a} & \textit{n/a} \\
    \textsc{wine} & 18.4 & $-3.2$ & $-0.7$$^{\dagger}$ & \textit{n/a} & \textit{n/a} \\
    \textsc{markdown} & 17.9 & $+4.5$ & $+5.5$ & $+1681.2$ & $+5.0$ \\
    \textsc{jsym} & 11.0 & $-1.9$ & $+2.0$ & \textit{n/a} & \textit{n/a} \\
    \textsc{lightgbm} & 7.0 & $+9.1$$^{\dagger}$ & $+7.7$$^{\dagger}$ & \textit{n/a} & \textit{n/a} \\
    \textsc{igraph} & 6.5 & $+2.3$$^{\dagger}$ & $+2.5$$^{\dagger}$ & $+16.0$$^{\dagger}$ & $+6.3$$^{\dagger}$ \\
    \textsc{compression} & 5.9 & $+0.6$$^{\dagger}$ & $-2.7$ & $+66.4$ & $+3.9$ \\
    \textsc{spacy} & 5.7 & $-3.5$ & $+3.1$ & \textit{n/a} & \textit{n/a} \\
    \textsc{dna-visualization} & 2.8 & $+7.9$ & $-0.9$$^{\dagger}$ & $-3.9$ & \textit{n/a} \\
    \textsc{textblob} & 2.6 & $-1.2$$^{\dagger}$ & $+1.0$$^{\dagger}$ & \textit{n/a} & \textit{n/a} \\
    \textsc{110.dynamic-html} & 1.4 & $+1.0$$^{\dagger}$ & $-1.7$ & $+158.6$ & \textit{n/a} \\
    \textsc{tensorflow} & 0.4 & $-5.1$ & $+2.9$$^{\dagger}$ & \textit{n/a} & \textit{n/a} \\
    \textsc{sentiment-gzip} & 0.3 & $+1.1$ & $+0.0$$^{\dagger}$ & \textit{n/a} & \textit{n/a} \\
    \textsc{shapely-numpy} & 0.1 & $+1.1$$^{\dagger}$ & $+5.0$ & $+1809.6$ & \textit{n/a} \\
    \textsc{scikit} & 0.1 & $-3.9$ & $+1.9$$^{\dagger}$ & \textit{n/a} & \textit{n/a} \\
    \textsc{encrypt} & 0.0 & $-9.6$ & $+0.0$$^{\dagger}$ & $+1159.6$ & $+15.5$ \\
    \midrule
    \textbf{Median} &  & $-1.2$ & $+0.6$ & $+41.2$ & $+5.0$ \\
    \textit{applications} &  & \textit{31} & \textit{28} & \textit{10} & \textit{7} \\
    \bottomrule
  \end{tabular}
\end{table}

\begin{table*}[!t]
  \caption{FaaSLight and \slimstart\ on AWS Lambda,
  on the 11 applications
  where either of them produced an artifact.
  The first column of each group is the median value
  of the unmodified application,
  and the tool columns give the change against it,
  where a negative change is faster or smaller.
  A change marked with $\dagger$ is not statistically significant,
  and for peak memory not above the 2\,MB reporting step.
  \textit{n/a} marks an application
  for which the tool produced no artifact.}
  \label{tab:appendix-secondary}
  \centering
  \small
  \setlength{\tabcolsep}{3.5pt}
  \begin{tabular}{@{}l rrr rrr rrr@{}}
    \toprule
    & \multicolumn{3}{c}{\textbf{Cold start (ms)}} & \multicolumn{3}{c}{\textbf{Initialization (ms)}} & \multicolumn{3}{c}{\textbf{Peak memory (MB)}} \\
    \cmidrule(lr){2-4} \cmidrule(lr){5-7} \cmidrule(l){8-10}
    \textbf{Application} & original & FaaSLight & \slimstart & original & FaaSLight & \slimstart & original & FaaSLight & \slimstart \\
    \midrule
    \textsc{ffmpeg} & 2{,}546 & $+3.2$\% & $+0.8$\%$^{\dagger}$ & 205 & $+9.2$\% & $+7.6$\% & 291 & $+0.7$\% & $+0.0$\%$^{\dagger}$ \\
    \textsc{chdb-olap} & 1{,}170 & $+1.5$\% & $-2.2$\%$^{\dagger}$ & 1{,}141 & $+1.6$\% & $-2.3$\%$^{\dagger}$ & 304 & $+0.3$\%$^{\dagger}$ & $+0.0$\%$^{\dagger}$ \\
    \textsc{lxml} & 589 & \textit{n/a} & $+2.8$\%$^{\dagger}$ & 356 & \textit{n/a} & $+1.9$\%$^{\dagger}$ & 71 & \textit{n/a} & $+0.0$\%$^{\dagger}$ \\
    \textsc{face-detection} & 536 & $+105.8$\% & \textit{n/a} & 387 & $+147.6$\% & \textit{n/a} & 97 & $+3.1$\% & \textit{n/a} \\
    \textsc{dna-visualization} & 370 & $+23.8$\% & \textit{n/a} & 352 & $+25.0$\% & \textit{n/a} & 68 & $+4.4$\% & \textit{n/a} \\
    \textsc{shapely-numpy} & 305 & $+30.8$\% & \textit{n/a} & 301 & $+30.4$\% & \textit{n/a} & 61 & $+4.9$\% & \textit{n/a} \\
    \textsc{110.dynamic-html} & 221 & $+17.5$\% & \textit{n/a} & 217 & $+16.5$\% & \textit{n/a} & 50 & $+2.0$\%$^{\dagger}$ & \textit{n/a} \\
    \textsc{igraph} & 209 & $+15.5$\% & $+0.3$\%$^{\dagger}$ & 205 & $+15.9$\% & $+0.4$\%$^{\dagger}$ & 47 & $+2.1$\%$^{\dagger}$ & $+0.0$\%$^{\dagger}$ \\
    \textsc{markdown} & 176 & $+144.5$\% & $+2.2$\%$^{\dagger}$ & 155 & $+12.9$\% & $+3.0$\%$^{\dagger}$ & 40 & $+5.0$\% & $+0.0$\%$^{\dagger}$ \\
    \textsc{encrypt} & 174 & $+8.2$\% & $+1.3$\%$^{\dagger}$ & 160 & $+0.4$\%$^{\dagger}$ & $-15.5$\% & 46 & $+2.2$\%$^{\dagger}$ & $+0.0$\%$^{\dagger}$ \\
    \textsc{compression} & 164 & $+3.3$\% & $+4.8$\% & 153 & $-3.1$\% & $+4.7$\% & 43 & $+0.0$\%$^{\dagger}$ & $+0.0$\%$^{\dagger}$ \\
    \midrule
    \textbf{Median} &  & $+16.5$\% & $+1.3$\% &  & $+14.4$\% & $+1.9$\% &  & $+2.2$\% & $+0.0$\% \\
    \bottomrule
  \end{tabular}
\end{table*}

\section{Results for FaaSLight and SlimStart}
\label{appendix:faaslight-slimstart-cold}

Table~\ref{tab:appendix-secondary} reports
FaaSLight and \slimstart\ on AWS Lambda,
on the 11 applications
where either of them produced an artifact,
10 for FaaSLight and 7 for \slimstart\
(Section~\ref{sec:eval-time}).
This subset is the only ground
on which the two can be measured at all,
since neither handles
the remaining 20 applications.
We omit their per-application comparison
against \xtrim\ and \ltrim,
which Table~\ref{tab:rq2} already gives,
and report here only the medians
the four tools reach on this subset.
\xtrim\ takes 5.4\% off the cold start
and \ltrim\ 4.1\%,
against an increase of 16.5\% under FaaSLight
and 1.3\% under \slimstart.
For initialization the four medians are
$-11.1$\%, $-10.4$\%, $+14.4$\% and $+1.9$\%,
and for peak memory
$-4.2$\%, $-2.3$\%, $+2.2$\% and $+0.0$\%.
Nine of the 11 are among
the twelve cheapest applications of the dataset
by unmodified cold start,
which is why \xtrim's median on the subset
is well below the 21.7\% it reaches
over all 31 applications
(Section~\ref{sec:eval-effectiveness}).
Section~\ref{sec:background} explains
why FaaSLight is not beneficial.

\end{document}